\documentclass[12pt,twoside,a4paper]{article}
\usepackage[tmargin=2cm, lmargin=2cm, bmargin=3cm,hmarginratio=1:1]{geometry} 
\usepackage[T1]{fontenc}
\usepackage[french,english]{babel}
\usepackage{color,graphicx,caption}
\usepackage{amsmath,amssymb,amsfonts}
\usepackage{amsthm}
\usepackage{mathtools}
\usepackage{sidecap} 
\usepackage{array}
\usepackage[indentafter]{titlesec}
\usepackage[utf8]{inputenc}
\usepackage{setspace}
\usepackage{perpage} 
\usepackage{hyperref} 

\MakePerPage{footnote} 
\newcommand{\p}{\partial}
\newcommand{\rd}{\mathcal}

\newcommand{\be}{\begin{eqnarray}}
\newcommand{\en}{\end{eqnarray}}
\newcommand{\badat}{\begin{alignedat}}
\newcommand{\eadat}{\end{alignedat}}
\newcommand{\bitm}{\begin{itemize}}
\newcommand{\eitm}{\end{itemize}}
\newcommand{\bmat}{\begin{pmatrix}}
\newcommand{\emat}{\end{pmatrix}}
\newcommand{\ba}{\begin{align}}
\newcommand{\bas}{\begin{align*}}
\newcommand{\ab}{\end{align}}
\newcommand{\bse}{\begin{subequations}}
\newcommand{\ese}{\end{subequations}}
\newcommand{\gt}{\rightarrow}
\newcommand{\ssi}{\Leftrightarrow}
\newcommand{\om}{\omega}
\newcommand{\ep}{\epsilon}
\newcommand{\virg}{\hspace{1 mm}, \hspace{8 mm}}
\newcommand{\pvirg}{\hspace{1 mm}; \hspace{8 mm}}

\newcommand{\e}{\hspace{2 mm}}
\newcommand{\ee}{\hspace{5 mm}}
\newcommand{\zb}{\bar z}

\numberwithin{equation}{section} 
\sidecaptionvpos{figure}{c} 

\begin{document}

\thispagestyle{empty}

\rightline{\small}

\vskip 3cm
\noindent
\begin{spacing}{1.5}
\noindent
{\Large \bf  Asymptotic dynamics of three-dimensional gravity}
 \end{spacing}
\vskip .6cm

\noindent
\linethickness{.03cm}
\line(1,0){430}
\vskip 1cm
\noindent
{\large \bf Laura Donnay\footnote{ldonnay@ulb.ac.be} }

\vskip 0.6cm

\noindent{\em 
Universit\'e Libre de Bruxelles and International  Solvay
Institutes}
\vskip 0.15cm
\noindent{\em
ULB-Campus Plaine CP231}
\vskip 0.15cm
\noindent{\em \hskip -.05cm B-1050 Brussels, Belgium}
\vskip1cm

\vskip 1cm

\vskip0.6cm

\noindent {\sc Abstract:} 
These are the lectures notes of the course given at the Eleventh Modave Summer School in Mathematical Physics, 2015, aimed at PhD candidates and junior researchers in theoretical physics. We review in details the result of Coussaert-Henneaux-van Driel showing that the asymptotic dynamics of $(2+1)$- dimensional gravity with negative cosmological constant is described at the classical level by Liouville theory. Boundary conditions implement the asymptotic reduction in two steps: the first set reduces the $SL(2,\mathbb R)\times SL(2,\mathbb R)$ Chern-Simons action, equivalent to the Einstein action, to a non-chiral $SL(2,\mathbb R)$ Wess-Zumino-Witten model, while the second set imposes constraints on the WZW currents that reduce further the action to Liouville theory. We discuss the issues of considering the latter as an effective description of the dual conformal field theory describing AdS$_3$ gravity beyond the semi-classical regime. 

\newpage
\tableofcontents

\newpage
\section{Introduction}

Three-dimensional general relativity provides us with an interesting toy model to investigate diverse aspects of gravity \cite{Deser1,Deser2} which, otherwise, would lie beyond our current understanding. While its dynamics is substantially simpler than the one of its four-dimensional analog, three-dimensional Einstein gravity still exhibits several phenomena that are present in higher dimensions and are still poorly understood, such as black hole thermodynamics. Remarkably, Einstein gravity in $2+1$ spacetime dimensions admits black hole solutions \cite{BTZ} whose properties resemble very much those of the four-dimensional black holes, as for instance the fact of having an entropy obeying the Bekenstein-Hawking area law. In order to admit theses black hole solutions in three-dimensions, the theory requires the presence of a negative cosmological constant \cite{Ida}, making the solutions to behave asymptotically as Anti-de Sitter (AdS) space. This makes this model even more powerful since it allows us to make use of AdS/CFT correspondence \cite{Maldacena} to ask fundamental questions about quantum gravity. Although a fully satisfactory quantum version of three-dimensional general relativity has not yet been accomplished \cite{MaloneyWitten}, this program has already shown to be promising, specially in the context of black hole physics \cite{Strominger}. 

The AdS/CFT correspondence permits to formulate questions about quantum gravity in AdS$_3$ spaces in terms of its much better understood dual two-dimensional conformal field theory (CFT). As a matter of fact, in three dimensions, the relation AdS$_3$/CFT$_2$ was discovered a long time before the holographic correspondence was formulated: In the work \cite{BrownHenneaux}, which, according to Witten, can be considered as the precursor of AdS/CFT correspondence, Brown and Henneaux showed that the symmetry algebra of asymptotically AdS$_3$ spaces is generated by two copies of Virasoro algebra with non-vanishing central charge, namely the algebra of local conformal transformations in two dimensions. At that moment, the appearance of a central charge in a classical context was surprising, while it is now understood as a crucial ingredient in the holographic context. Soon later, it was shown \cite{CHvD} that the asymptotic dynamics of Einstein gravity around AdS$_3$ space is governed by the Liouville action, a non-trivial two-dimensional conformal field theory whose central charge coincides with the one found in \cite{BrownHenneaux}. These notes are aimed at describing the connection existing between these theories, showing in detail all the steps that brings one from three-dimensional Einstein gravity in asymptotically AdS spaces to the Liouville field theory action. Finding the dual conformal field theory living at the boundary of three-dimensional gravity would solve quantum gravity in three-dimensions; the classical computation we will present shows that the dual theory has to admit Liouville theory as an effective description in a certain regime. \\

These notes are organized as follows: After reviewing in Section \ref{sec3D} the action of gravity in $2+1$ spacetime dimensions and the absence of local degrees of freedom, we present in Section \ref{secBTZ} the black hole solution hosted in the case of negative cosmological constant. In Section \ref{secCS}, we show how the Einstein-Hilbert action can be written as a Chern-Simons action for the gauge group $SL(2,\mathbb{R}) \times SL(2,\mathbb{R})$, using the vielbein and spin connection formalism. In Section \ref{secBC}, we present the Brown-Henneaux AdS$_3$ boundary conditions and compute the associated asymptotic symmetry algebra in the Chern-Simons formalism. Section \ref{secWZW} contains a brief introduction on Wess-Zumino-Witten (WZW) models. The two steps of the reduction of the Chern-Simons action to, first, a non-chiral WZW model, and then to a Liouville action, are detailed in Sections \ref{secCSWZW} and \ref{secWZWLiouv}. After a brief introduction on Liouville theory, in Section \ref{secLiouv}, we discuss in Section \ref{secEnt} the possibility of the latter to account for the microstates of the BTZ black hole.

\section{Gravity in \texorpdfstring{$2+1$}{2p1} dimensions}
\label{sec3D}
Pure gravity in $2+1$ spacetime dimensions is defined by the three-dimensional Einstein-Hilbert action (where we set the speed of light to $c \equiv 1$):
\be
S_{\mathrm{EH}}[g]\equiv \frac{1}{16 \pi G} \int_{\rd M} \, d^3x \, \sqrt{-g} \, (R-2 \Lambda) + B,
\label{EHaction}
\en
with $G$ the three-dimensional Newton constant, $g\equiv \mathrm{det} g_{\mu \nu}$ ($\mu, \nu=0,1,2$), with a metric $g_{\mu \nu}$ of signature $(-,+,+)$, $R\equiv R_{\mu \nu}g^{\mu \nu}$ is the curvature scalar, and $R_{\mu \nu}$ the Ricci tensor. $\rd M$ is a three-dimensional manifold, and $\Lambda$ is the cosmological constant, which can be positive, negative, or null, yielding respectively locally de Sitter (dS), Anti-de Sitter (AdS), or flat spacetimes. Here, we will be concerned with Anti-de Sitter spacetime, for which the cosmological constant is related to the AdS radius $\ell$ through $\Lambda=-1/\ell^2$. Action \eqref{EHaction} is defined up to a boundary term $B$, which is there in order to ensure that the action has a well-defined action principle.

Extremizing the action with respect to the metric $g_{\mu \nu}$ yields the Einstein equations
\be
R_{\mu \nu}-\frac{1}{2}g_{\mu \nu} R+\Lambda g_{\mu \nu}=0.
\label{Eeq}
\en
An important property of general relativity in $2+1$ dimensions is that any solution of the vacuum Einstein equations \eqref{Eeq} with $\Lambda <0$ is locally Anti-de Sitter (locally de Sitter if $\Lambda > 0$ and locally flat if $\Lambda=0$). This can be verified by realizing that the full curvature tensor in three dimensions is totally determined by the Ricci tensor\footnote{In three dimensions, the Weyl tensor vanishes identically.}
\be
R_{\mu \nu \rho \sigma}=g_{\mu \rho}R_{\nu \sigma}+g_{\nu \sigma}R_{\mu \rho}-g_{\nu \rho}R_{\mu \sigma}-g_{\mu \sigma}R_{\nu \rho}-\frac{1}{2}R(g_{\mu \rho}g_{\nu \sigma}-g_{\mu \sigma}g_{\nu \rho}).
\en
As a consequence, any solution of Einstein equations \eqref{Eeq} has constant curvature; namely
\be
R_{\mu \nu \rho \sigma}=\Lambda(g_{\mu \rho}g_{\nu \sigma}-g_{\mu \sigma}g_{\nu \rho}).
\label{tenseur}
\en

Physically, this means that on three-dimensional Einstein spacetimes there are no local propagating degrees of freedom: there are no gravitational waves in this theory. Another way to see that gravity in $D=3$ dimensions has no degrees of freedom (d.o.f.) is counting them explicitly: Out of the $D(D+1)/2$ components of a symmetric tensor $g_{\mu \nu}$ in $D$ spacetime dimensions, one can always remove $D$ of them using diffeomorphism invariance (one removes one d.o.f. per coordinate). Moreover, $D$ components of the metric appear in the Lagrangian with no temporal derivative, they are therefore no true d.o.f. but Lagrange multipliers. This counting leads therefore to
\be
\frac{D(D+1)}{2}-D-D =0 \ee (D=3).
\en 
A priori, this property may look very disappointing: How could this theory be a realistic model to study four-dimensional gravity if there is no graviton at all? However, despite the absence of local d.o.f., it turns out that three-dimensional gravity is in fact far from being sterile. This is because of two fundamental reasons: first, as we will see later, even though every spacetime is locally equivalent to a constant curvature spacetime, it may differ from the maximally symmetric solution by \emph{global} properties, and this allows for interesting geometrical properties such as non-trivial causal structures. The second reason is that, unexpectedly, in the case $\Lambda<0$, there exist black hole solutions.


\section{The three-dimensional black hole}
\label{secBTZ}

To everyone's surprise, Bañados, Teitelboim and Zanelli (BTZ) showed in 1992 that $2+1$-dimensional gravity admits a black hole solution \cite{BTZ} that shares many physical properties with the four-dimensional Kerr black hole. 
The BTZ black hole\footnote{For a review, see \cite{CarlipBTZ}.} of mass $M$ and angular momentum $J$ is described, in Schwarzschild type coordinates, by the metric
\be
\label{BTZ}
ds^2=-\left(N(r)\right)^2dt^2+\left(N(r)\right)^{-2}dr^2+r^2\left(d\varphi+N^{\varphi}(r)dt\right)^2,
\en
where the lapse and shift functions are given by\footnote{Notice that many authors set $8G \equiv1$.}
\be
\label{lapseShift}
N(r)\equiv\sqrt{-8GM+\frac{r^2}{\ell^2}+\frac{16G^2J^2}{r^2}}\quad\mathrm{,}\quad N^{\varphi}(r)\equiv-\frac{4GJ}{r^2}.
\en
with $-\infty <t<+\infty$, $0<r<+\infty$ and $0 \leq \varphi \leq 2\pi$. 
It solves the Einstein equation \eqref{Eeq} with cosmological constant $\Lambda=-1/\ell^2$.

The BTZ metric (\ref{BTZ}) is stationary and axially symmetric, with Killing vectors $\partial_ t$ and $\partial_\varphi$. This metric exhibits a (removable) singularity at the points $r=r_\pm$ where $N(r_{\pm })=0$; that is,
\be
r_{\pm}=\ell\left[4G M\left(1\pm\sqrt{1-\left(J/M \ell\right)^2}\right)\right]^{1/2}.
\en
When $|J|\leq M \ell$, the BTZ possesses an event horizon at $r_+$ and an inner Cauchy horizon (when $J \neq 0$) at $r_-$. In terms of $r_{\pm }$, the mass and angular momentum read
\be
M=\frac{r_+^2+r_-^2}{8G \ell^2} \virg J=\frac{r_+r_-}{4G\ell}.
\label{MJ}
\en

\begin{SCfigure}
\centering
\includegraphics[width=0.5 \textwidth]{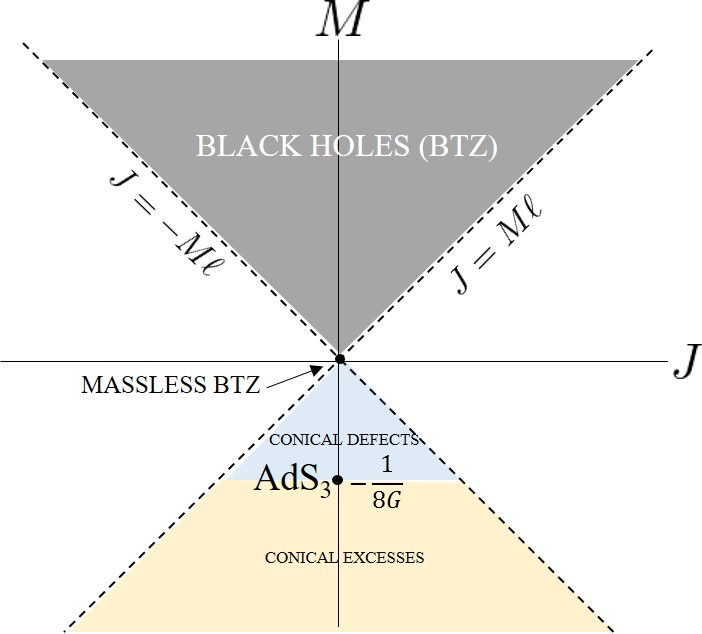}
\captionsetup{width=1\linewidth}
\caption{Spectrum of the BTZ black hole. Black holes exist for $M \geq 0$, $|J|\leq M \ell$. The vacuum state $M=-\frac{1}{8G}$, $J=0$, separated by a gap from the continuous spectrum, corresponds to AdS$_3$. }
\label{BTZim}
\end{SCfigure}

In the case $|J|=M\ell$, both horizons coincide $r_+=r_-$; this case corresponds to the so-called extremal BTZ. If $M<0$ (or if $|J|$ becomes too large), the horizon at $r=r_+$ disappears, leading therefore to a naked singularity at $r=0$. Relation $|J|\leq M \ell$ plays therefore the role of a cosmic censorship condition. There is, however, a special case: when $M=-1/(8G)$ and $J=0$, both the horizon and the singularity disappear! At this point, the metric exactly coincides with (the universal covering of) AdS$_3$ spacetime; namely
\[
ds^2_{M=-\frac{1}{8G}, \,J=0}=-\left(1+\frac{r^2}{\ell^2}\right)dt^2+\left(1+\frac{r^2}{\ell^2}\right)^{-1}dr^2+r^2d\varphi^2.
\]
Therefore, AdS spacetime is separated from the continuous spectrum of the BTZ black holes by a mass gap of $\Delta_0 = 1/(8G)$, see Figure \ref{BTZim}. The solution with $-1/(8G) < M <0$ corresponds to naked singularities, with conical singularity at the origin. These solutions exhibit an angular deficit around $r=0$ and admit to be interpreted as particle-like objects \cite{Deser1}. Therefore, one cannot continuously deform a black hole state to the AdS$_3$ vacuum, since it would imply to go through the regions with naked singularities. The solutions with $M<-1/(8G)$ also correspond to naked singularities, in this case with angular excesses around $r=0$.

Since, as we saw above, any solution of pure gravity in three dimensions is locally of constant curvature, the BTZ solution \eqref{BTZ} is locally Anti-de Sitter: every point of the black hole has a neighborhood isometric to AdS$_3$ spacetime, and therefore the whole black hole can be expressed as a collection of patches of AdS assembled in the right way. The fact that BTZ differs from AdS only by global properties suggests that the black hole metric can be obtained by identifying points of AdS spacetime by a sub-group of its isometry group. That is actually what Henneaux, Bañados, Teiteilboim and Zanelli showed in \cite{HBTZ}, where these identifications were given explicitly. 

Very far away from the black hole, namely when $r\gg r_+$, the metric reduces to 
\be
ds^2_{M=J=0}=-\frac{r^2}{\ell^2}dt^2+\frac{\ell^2}{r^2}dr^2+r^2d\varphi^2.
\label{vac}
\en
The asymptotic behavior\footnote{By asymptotic, here simply mean far away. We will give a more precise definition in Section \ref{secBC}.} of the BTZ black hole and AdS$_3$ is thus the same, this is why the BTZ is said to be asymptotically AdS. This is in contrast with the Schwarzschild and Kerr black holes which are asymptotically flat.  In fact, there is no black hole asymptotically flat, nor asymptotically de Sitter in three-dimensions (for pure gravity) \cite{Ida}.

The $g_{00}$ component of the BTZ is zero at $r=r_{\mathrm{erg}}$, with
\be
r_{\mathrm{erg}}=\sqrt{r_+^2+ r_-^2}=\ell \sqrt{8G M}.
\label{erf}
\en
The $r<r_{\mathrm{erg}}$ region is called \emph{ergosphere}, meaning that all observers in this region are unavoidably dragged along by the rotation of the black hole. The existence of an event horizon and of an ergosphere region make the BTZ extremely similar to the four-dimensional Kerr black hole. Another feature that BTZ shares with the Kerr solution is that, in both spaces, the surface $r=r_-$ is a Killing horizon. In fact, even though BTZ has no curvature singularity at the origin, it is quite similar to realistic (3+1)-dimensional black holes: it is the final state of gravitational collapse \cite{Ross}, and possesses similar properties also at quantum level. Indeed, remarkably, the BTZ exhibits non-trivial thermodynamical properties; it radiates at a Hawking temperature
\be
T_{\text{BH}}=\frac{\hbar (r_+^2-r_-^2)}{2\pi \ell^2 r_+},
\en
and has a Bekenstein-Hawking entropy
\be
S_{\text{BH}}=\frac{A}{4 \hbar G},
\label{EntBH}
\en
with $A=2\pi r_+$ the horizon size. That is, three-dimensional black holes also obey the {\it area} law, whose full microscopic understanding is one of the main questions in quantum gravity. Besides, these thermodynamical quantities satisfy the first law of black hole thermodynamics 
\begin{equation}
dM=T_{\text{BH}} dS_{\text{BH}}+\Omega \ dJ,
\end{equation} 
where $\Omega=r_-/(r_+ \ell)$ is the angular velocity at the horizon. Notice that the BTZ also exhibits a Hawking-Page transition at $r_+ \sim \ell$. 

Finally, it is worth mentioning that, besides pure gravity, the BTZ solution appears in many other frameworks, such as supergravity \cite{Couss}, string theories \cite{Horo}, and higher-spins \cite{Kraus}. Moreover, the BTZ turned out to appear in the near horizon limit of higher-dimensional solutions \cite{MS}; all of this showing the relevance of this black hole solution in more general set-ups.


\section{\texorpdfstring{$3D$}{3D} gravity as a gauge theory} 
\label{secCS}
A crucial property of three-dimensional gravity action \eqref{EHaction} is that it can be rewritten in terms of ordinary gauge fields, in such a way that both the structure of the action and equations of motion simplify substantially. This fact was discovered by Achúcarro and Townsend \cite{AT}, and latter by Witten \cite{Witten1}, and holds for any sign of the cosmological constant. The validity of this result can be extended to supergravity actions \cite{AT,Banados1}, as well as higher-spins \cite{CSHS1,CSHS2}. 

\subsection{Vielbein and spin connection formalism}

This result is based on the \emph{first-order}, or Palatini formulation of general relativity. This consists in the following: Instead of working as we usually do with the metric $g_{\mu \nu}$, we will use an auxiliary quantity $e^a_\mu$ (with a frame index $a=0,1,2$), called \emph{frame field}, or \textit{vielbein}\footnote{In three dimensions, it receives the name {\it dreibein}; {\it vierbein} or {\it tetrad} in four dimensions.}, which can be thought of as the square root of the metric\footnote{Its existence comes from the fact that the metric tensor can be diagonalized by an orthogonal matrix $O^a_\mu$ with positive eingeinvalue $\lambda^a$, the vielbein is therefore defined as $e^a_\mu=\sqrt{\lambda^a}O^a_\mu$.}; namely
\be
g_{\mu \nu}(x)=e^a_\mu (x) \eta_{ab}e^b_\nu(x),
\label{viel}
\en
where $\eta_{ab}$ is the metric of flat $3D$ Minkowski spacetime.

Relation \eqref{viel} can be simply seen as the transformation of a tensor under a change of coordinates described by the matrix $e^a_\mu$. Since $e^a_\mu$ is a non-singular matrix, with $e\equiv \mathrm{det }e^a_\mu=\sqrt{- \mathrm{det}g}\neq 0$, there is an inverse frame field $e^\mu_a(x)$ such that $e^a_\mu e^\mu_b=\delta^a_b$ and $e^\mu_a e^a_\nu= \delta^\mu_\nu$.
Notice that, for a given metric, the frame field is not unique; indeed, all frame fields related by a local Lorentz transformation $e'^a_\mu=\Lambda^{-1 a}_{\e b}(x)e^b_\nu (x)$ with $\Lambda \in SO(2,1)$ are equivalent (the transformation is local since it affects only the frame indices, while the spacetime indices do not see such transformation).

We can use the vielbein to define a basis in the space of differential forms. We define the one-form $e^a\equiv e^a_\mu dx^\mu$ and the Levi-Civita in frame components $\ep_{abc}$ in the following way:
\be
\badat{2}
&\ep_{\mu \nu \rho}\equiv e^{-1} \ep_{abc}e^a_\mu e^b_\nu e^c_\rho,\\
&\ep^{\mu \nu \rho}\equiv e\, \ep^{abc}e_a^\mu e_b^\nu e_c^\rho.
\label{epsi}
\eadat
\en

A covariant derivative is made out of the normal derivative plus an affine connection; schematically $D=\p + \Gamma$.
In the tetrad formalism, the role of this connection is played by one-forms $\om^{ab}=\om^{ab}_\mu dx^\mu$, with $\om^{ab}=-\om^{ba}$. This quantity is introduced because it permits to construct a quantity that transforms as a local Lorentz vector. Indeed, unlike the 2-form $de^a$, the following quantity, called the \emph{torsion} 2-form of the connection,
\be
T^a \equiv d e^a+\om^a_{\e b}\wedge e^b \ ,
\label{torsion}
\en
does transform as a vector under local Lorentz transformation, namely $T^a \gt \Lambda^{-1a}_{\ee b}T^b$, provided the quantity $\om^a_{\e b}$, whose components $\om_\mu^{ab}$ are called \emph{spin connections}, transforms as
\be
\om^{ a}_{\e b} \gt \Lambda^{-1a}_{\ee c}d\Lambda^c_{\e b}+\Lambda^{-1a}_{\ee c}\om^c_{\e d}\Lambda^d_{\e b}.
\en
Equation \eqref{torsion} is called the first Cartan structure equation. The second Cartan structure equation is given by
\be
d\om^{ab}+\om^a_{\e c}\wedge \om^{cb}=R^{ab},
\en
with 
\be
R_{\mu \nu}^{ab}(\om)=\partial_{\mu} \omega_{\nu}^{ab}-\partial_{\nu} \omega_{\mu}^{ab}+\omega_\mu^{ac}\omega_{\nu c}^{b}-\omega_\nu^{ac}\omega_{\mu c}^{b}
\en
the curvature tensor, and
\be
\badat{2}
&R^{ab}=\frac{1}{2}R_{\mu \nu}^{ab}(x) dx^\mu \wedge dx^\nu,\\
&R^{\lambda \sigma}_{\e \mu \nu}=e^\lambda_a e^\sigma_b R^{ab}_{\e \mu \nu}.
\label{RR}
\eadat
\en

We can of course relate the spin connection with the usual Christofell symbols; the relation is $\Gamma^\rho_{\mu \nu}=e^\rho_a(\partial_\mu e^a_\nu+\om_{\mu \e b}^{\e a}e^b_\nu)$. To see that the spin connection $\om$ plays the role of a gauge field, notice the analogy with the Yang-Mills connection $A$:
\be
\badat{2}
&D=\p +A \virg F=dA+A\wedge A,\\
&D=\p +\om \virg R=d\om+\om\wedge \om.
\eadat
\en

Let us now go back to our three-dimensional Einstein-Hilbert action. In terms of the quantities we have defined above, \eqref{EHaction} reads (we will take care of the boundary term later)
\be
S_{\text{EH}}[e,\om]=\frac{1}{16 \pi G} \int_{\rd M} \, \ep_{abc}\left(e^a \wedge R^{bc}[\om]-\frac{\Lambda }{3} e^a \wedge e^b \wedge e^c \right).
\label{EH1}
\en
Indeed, using \eqref{epsi}, we notice that
\be
\badat{2}
d^3x \sqrt{-g}&=e \,dx^0 dx^1 dx^2=\frac{1}{3!}e \,\ep_{\mu \nu \rho}dx^\mu \wedge dx^\nu \wedge dx^\rho\\
&=\frac{1}{3!}\ep_{abc}e^a \wedge e^b \wedge e^c;
\eadat
\en
and, using \eqref{RR}, we have
\be
\badat{2}
\ep_{abc}\,e^a \wedge R^{bc}
&=\frac{1}{2}  e \,\ep_{\mu \alpha \beta}  R^{\alpha \beta}_{\e \nu \rho}\ep^{\mu \nu \rho} d^3x \\
&=d^3x \sqrt{-g}R.
\eadat
\en
In what follows, we will adopt the so-called {\it dual} notation (valid only in three dimensions),
\be
\badat{2}
&R_a \equiv \frac{1}{2}\ep_{abc}R^{bc} \, \leftrightarrow R^{ab}\equiv -\ep^{abc}R_c, \\
&\om_a \equiv \frac{1}{2}\ep_{abc}\om^{bc} \, \leftrightarrow \om^{ab}\equiv -\ep^{abc}\om_c.
\eadat
\en
With this, we notice that the gravity action \eqref{EH1} can finally be rewritten as
\be
S_{\text{EH}}[e,\om]=\frac{1}{16 \pi G} \int_{\rd M} \, \left(2 \,e^a \wedge R_a[\om]-\frac{\Lambda }{3} \ep_{abc} e^a \wedge e^b \wedge e^c \right).
\label{EH2}
\en

\subsection{The Chern-Simons action}
Now that we have at hand the gravity action in terms of the vielbein and the spin connection, we are ready to prove, as announced above, that three-dimensional gravity is equivalent to a gauge theory with a specific kind of interaction, called Chern-Simons theory. We will first introduce this very interesting model for the readers who are not familiar with it. For an introduction to Chern-Simons in $3D$, see for instance \cite{Blago}.

A \textit{Chern-Simons action} for a compact gauge group $G$ is given by
\be
S_{\mathrm{CS}}[A]=\frac{k}{4 \pi} \int_{\rd M} \mathrm{Tr} \left[A \wedge dA +\frac{2}{3}A\wedge A\wedge A \right],
\label{CS1}
\en
where $k$ is a constant called \textit{level}, while the gauge field $A$ represents a Lie algebra-valued one-form $A=A_\mu dx^\mu$, and Tr represents a non-degenerate\footnote{This is asked in order that all gauge fields have a kinetic term in the action; this is always true for semisimple Lie algebras.} invariant bilinear form on the Lie algebra (of the gauge group $G$).

Integrating by parts, the variation of action \eqref{CS1} takes the form
\be
\delta S_{CS}[A]=\frac{k}{4 \pi} \int_{\rd M} \mathrm{Tr} \left[2 \delta A \wedge (dA +A\wedge A)  \right]-\frac{k}{4 \pi} \int_{\partial \rd M} \mathrm{Tr} \left[A \wedge  \delta A \right].
\en
If $\delta A$ is chosen such that its value on the boundary $\p \rd M$ is such that the second term vanishes, we obtain
\be
F\equiv dA+A \wedge A = 0,
\en
where $F$ is the usual field strength 2-form.
These equations imply that, locally, 
\be
A=G^{-1}dG,
\label{gauge}
\en
which means that $A$ is a gauge transformation of the trivial field configuration; in other words, $A$ is \textit{pure gauge}. Therefore, a Chern-Simons theory has no true propagating degrees of freedom: it is purely topological. Indeed, all the physical content of the theory is contained in non-trivial topologies, which prevent relation \eqref{gauge} to hold everywhere on the manifold $\rd M$. 

If we write $A=A^a T_a$, with $T_a$ a basis\footnote{This has nothing to do with the torsion $T^a$ mentioned in the previous subsection.} of the Lie algebra of the gauge group $G$, then one has, for the first term of \eqref{CS1},
\be
 \mathrm{Tr} \left[A \wedge dA\right]=  \mathrm{Tr}(T_a T_b) \left[A^a \wedge dA^b\right].
\en
We then see that $d_{ab}\equiv \mathrm{Tr}(T_a T_b)$ plays the role of a metric on the Lie algebra, and therefore should be non-degenerate. The existence of a Chern-Simons action, and the form it will take, relies on whether the gauge group one wants to consider admits such an invariant non-degenerate form. Notice that one can make use of this bilinear form to define an inner product $(\cdot,\cdot)$.

\subsection{\texorpdfstring{$\Lambda <0$}{Lambda} gravity as a Chern-Simons theory for \texorpdfstring{$SO(2,2)$}{so22}}
What Achúcarro, Townsend and Witten discovered \cite{AT,Witten1} is that three-dimensional gravity action and equations of motions are equivalent to a Chern-Simons theory for an appropriate gauge group. More precisely, their result states that pure gravity (Einstein-Hilbert action) is equivalent to a three-dimensional Chern-Simons theory based on the gauge group $SO(2,2)$ for $\Lambda <0$, $ISO(2,1)$ for $\Lambda = 0$, or $SO(3,1)$ for $\Lambda >0$. 

We will prove this result for the case $\Lambda <0$, since we are interested in Anti-de Sitter spaces. In this case, the Lie algebra involved is $so(2,2)$, whose commutation relations are given by
\be
[J_a, J_b]=\ep_{abc}J^c \virg [J_a, P_b]=\ep_{abc}P^c \virg [P_a, P_b]= \ep_{abc}J^c,
\label{alg}
\en
where the indices $a,b,c = 0,1,2$ are raised and lowered with the three-dimensional Minkowski metric $\eta_{ab}$ and its inverse $\eta^{ab}$. In (\ref{alg}), we have used the three-dimensional rewriting
\be
J_a \equiv \frac{1}{2}\ep_{abc}J^{bc} \, \leftrightarrow J^{ab}\equiv -\ep^{abc}J_c,
\en
where the $J_{ab}$ are the usual Lorentz generators, while the $P_a$ are the generators of the translations. This Lie algebra admits the following non-degenerate invariant (symmetric and real) bilinear form\footnote{This is not the only one, though; see the comments at the end of this section.}
\be
(J_a,P_b)=\eta_{ab} \virg (J_a,J_b)=0=(P_a,P_b).
\label{bform1}
\en
One then constructs the gauge field $A$ living on this Lie algebra as
\be
A_\mu\equiv \frac{1}{\ell }e^a_\mu P_a +\om^a_\mu J_a.
\label{A}
\en
Notice that the Lie algebra indices are identified with the frame indices of the vielbein and spin connection; this is crucial for the gravity $\leftrightarrow$ gauge theory relation that we are about to show. 
Equipped with the gauge field \eqref{A} and with the non-degenerate invariant form \eqref{bform1} one can write the Chern-Simons action \eqref{CS1} for the gauge group $G=SO(2,2)$. The first term is 
\be
\badat{3}
\mathrm{Tr}[A \wedge dA]&=(\frac{1}{\ell }e^a P_a +\om^a J_a,\frac{1}{\ell }de^b P_b +d\om^b J_b)\\
&=\frac{1}{\ell }\left(e^a  \wedge d\om^b +\om^a \wedge de^b \right)\eta_{ab}=\frac{2}{\ell }e^a  \wedge d\om_a,
\eadat
\en
while the second term is found to be
\be
\badat{3}
\frac{2}{3}\mathrm{Tr}[A \wedge A \wedge A]&=\frac{1}{3}\mathrm{Tr}[[A,A] \wedge A]\\
&=\frac{1}{3 \ell}\left(\frac{1}{\ell^2} e^a \wedge e^b \wedge e^c+3 \ep_{abc}e^a \wedge \om^b \wedge \om^c \right).
\eadat
\en
Therefore, we find that the Chern-Simons action for the group $SO(2,2)$ is equal to
\be
S_{\mathrm{CS}}[e,\om]=\frac{k}{4\pi \ell} \int_{\rd M} \, \left(2 e^a \wedge R_a[\om]+\frac{1}{3 \ell^2}\ep_{abc}\, e^a \wedge e^b \wedge e^c \right),
\en
where we have remembered that $R_a=d\om_a+\frac{1}{2}\ep_{abc}\om^b\wedge \om^c$. We have thus shown that the Chern-Simons action for $SO(2,2)$ exactly matches the Einstein-Hilbert action \eqref{EH2} with $\Lambda=-1/\ell^2$, provided that the level acquires the value\footnote{The sign of $k$ depends on the identity $\sqrt -g= \pm e$, namely depends on the choice of relative orientation of the coordinate basis and the frame basis.}
\be
k=\frac{\ell}{4 G}.
\en

The fact that Einstein gravity is merely a Chern-Simons action reminds us that there is no propagating degree of freedom in the theory, and thus no graviton in three-dimensions, since we saw above that this gauge theory is purely topological.
However, even though there are no local excitations, its dynamical content is far from being trivial due to the existence of boundary conditions\footnote{In other words, the relevant d.o.f. are \emph{global}. Introducing boundary conditions is not the only way to generate global d.o.f., one can also consider holonomies (we will not study them here). Notice however that holonomies generate only a finite number of degrees of freedom in the theory.}. We will see in Section \ref{secBC} that, under an appropriate choice of boundary conditions, there is in fact not one, not two, but an \textit{infinite number of degrees of freedom living on the boundary}. Boundary conditions are necessary in order to ensure that the action has a well-defined variational principle, but the choice of such conditions is not unique. In fact, the dynamical properties of the theory are extremely sensitive to the choice of boundary conditions. In this context, the residual gauge symmetry on the boundary is called global symmetry or \textit{asymptotic symmetry}. The breakdown of gauge invariance at the boundary has the effect of generating this infinite amount of degrees of freedom.\\

Before concluding this section, let us mention a very useful fact, namely the isomorphism $so(2,2) \approx sl(2,\mathbb{R})\oplus sl(2,\mathbb{R})$ (recall that $so(2,2)$ is semi-simple). 
Defining $
J_a^\pm \equiv \frac{1}{2}\left(J_a\pm P_a\right)$,
algebra \eqref{alg} reads
\be
[J_a^+, J_b^+]=\ep_{abc}J^{+c} \virg [J_a^-, J_b^-]=\ep_{abc}J^{-c} \virg [J_a^+, J_b^-]=0.
\en
Thanks to this splitting, one can rewrite the Chern-Simons action for the $so(2,2)$ connection\footnote{which we previously called $A$} $\Gamma$ as the sum of two Chern-Simons actions, each having their connections $A$, $\bar A$ in the first and second chiral copy of $sl(2,\mathbb R)$ respectively:
\be
A =(e^a/\ell +\om^a)T_a \virg \bar A =(e^a/\ell -\om^a)T_a,
\en
with $T_a$ now being the generators of $sl(2, \mathbb R)$. One can show that the decomposition of the action then reads
\be
S_{\mathrm{CS}}[\Gamma]=S_{\mathrm{CS}}[A]-S_{\mathrm{CS}}[\bar A] \equiv S_{\mathrm{CS}}[A, \bar A],
\en
that is, can be rewritten as the difference of a chiral and anti-chiral Chern-Simons action.

Finally, let us mention that Einstein's equations of motion are equivalent to the ones in the Chern-Simons formalism, namely $F^a=0$, $\bar F^a=0$.
More precisely, varying the action with respect to $e^a$ gives the constant curvature equation
\be
F^a+\bar F^a= 0 \, \ssi \, R^{ab}+\frac{1}{\ell^2}e^a \wedge e^b=0,
\en
while varying with respect to $\om^a$ leads to the torsion free equation
\be
F^a-\bar F^a= 0 \, \ssi \, T^a=de^a +\om^a_{\e b}\wedge e^b=0.
\en
We thus verify that solving the equations of motion in the Chern-Simons formalism is considerably simpler than solving Einstein's equations.

\subsection{Some comments on Chern-Simons theories} 

Before concluding this section, let us make some remarks on Chern-Simons theories that can be relevant for their gravity application. 

Let us begin by noticing that the Chern-Simons description of gravity is valid when the vielbein is invertible, which is true for classical solutions of gravity. However, from the gauge theory point of view, this is not entirely natural. This is relevant because, despite the identity between the actions of the two theories, it is not obvious that gravity and Chern-Simons are equivalent at quantum level where, besides the action, one has to provide a set of configurations over which to perform the functional sum. Perturbatively, close to classical saddle points, the relation between the gauge theory and three-dimensional gravity may remain valid; however, it is not clear that the relation still holds non-perturbatively \cite{Wittenrevisited}. Moreover, to claim that Chern-Simons and gravity theories are equivalent, we have to prove that the gauge transformations and the diffeomorphisms do match (up to a local Lorentz transformation). It is shown in \cite{Witten1} that this matching occurs only when the equations of motion are satisfied, namely on-shell.
	
A second comment regards the definition of the invariant bilinear form appearing in (\ref{CS1}). In addition to \eqref{bform1}, $so(2,2)$ admits a second one, given by 
\be
(J_a,J_b)=\eta_{ab} \virg (J_a,P_b)=0 \virg (P_a,P_b)= \eta_{ab},
\label{bform2}
\en
consequence of the isomorphism $so(2,2) \approx sl(2,\mathbb{R})\oplus sl(2,\mathbb{R})$. One can use this new form to construct an alternative Chern-Simons action: the so-called \textit{exotic} action \cite{Witten1}, which corresponds to
\be
S_{\mathrm{E}}[\Gamma]=S_{\mathrm{CS}}[A]+S_{\mathrm{CS}}[\bar A].
\en
This action is relevant in the construction of the so-called Topologically Massive Gravity \cite{TMG}.
	
Finally, let us mention an interesting feature of the Chern-Simons coupling constant, the level $k$. Generally, a Chern-Simons theory admits to be constructed by starting from a gauge invariant action $S_P$ defined on a four-dimensional manifold $\rd M_4$ whose boundary is the three-dimensional space $\rd M$ where the Chern-Simons theory is defined. Here, we think of a gauge field in four dimensions that is an extension of the three-dimensional gauge field of Chern-Simons theory. The extension of the three-dimensional gauge field is, generically, non-unique, and this carries information about topology. In fact, the four-dimensional action $S_P$ corresponds to a topological invariant; its Lagrangian density is a total derivative\footnote{A similar example is the Einstein-Hilbert action in two dimensions, which coincides with the Euler characteristic $\Xi=\frac{1}{2 \pi}\int_{\rd M_2} \sqrt{-g}R d^2x=2-2g$, where $g$ is the genus of the closed manifold ${\rd M_2}$.}. Topological invariant actions exist only in even dimension, and that is the reason why Chern-Simons actions exist only in odd dimensions. The four-dimensional action is given by
\be
S_P=\frac{k}{4 \pi} \int_{\rd M_4} \, \mathrm{Tr}(F \wedge F)=\frac{k}{4 \pi} \int_{\rd M_4} \, P,
\en
with $P\equiv d_{ab} F^a \wedge F^b$, $F$ being the curvature associated to the four-dimensional gauge field that extends the $A$ appearing in (\ref{CS1}). $P\equiv d_{ab} F^a \wedge F^b$ is called the Pontryagin form, and is a total derivative; more precisely $P=dL_{\mathrm{CS}}$, where $L_{\mathrm{CS}}$ is the Chern-Simons Lagrangian of (\ref{CS1}). Therefore, if $\partial \rd M_4=\rd M$, one can, after using Stokes' theorem, rewrite $S_P$ as an integral over $\rd M$, and one is left with the three-dimensional Chern-Simons action \eqref{CS1}.\\
Being a topological invariant, $S_P$ takes discrete values. In fact, one can show that $\int \mathrm{Tr}(F \wedge F)=4\pi^2 n$, with $n \in \mathbb Z$. This, together with asking that the action is defined modulo $2\pi$ (so that $e^{iS}$, which appears in the path integral, is single-valued), leads to the conclusion that, for the theory to be well defined, the Chern-Simons level has to be quantized, namely $k  \in \mathbb Z$ \cite{Wittenrevisited}.


\section{Asymptotically AdS\texorpdfstring{$_3$}{3} spacetimes}
\label{secBC}
In this section, we will make more precise what we mean by \emph{asymptotically Anti-de Sitter} spacetimes. We will consider a set of metrics which tend to the metric of AdS$_3$ in a specific way. Giving such information is actually equivalent to prescribing fall-off conditions on the metric components at large distances, the so-called \emph{boundary conditions}. Before that, we need to specify what is the boundary of our spacetime. 

\begin{SCfigure}
\centering
\includegraphics[width=0.2 \textwidth]{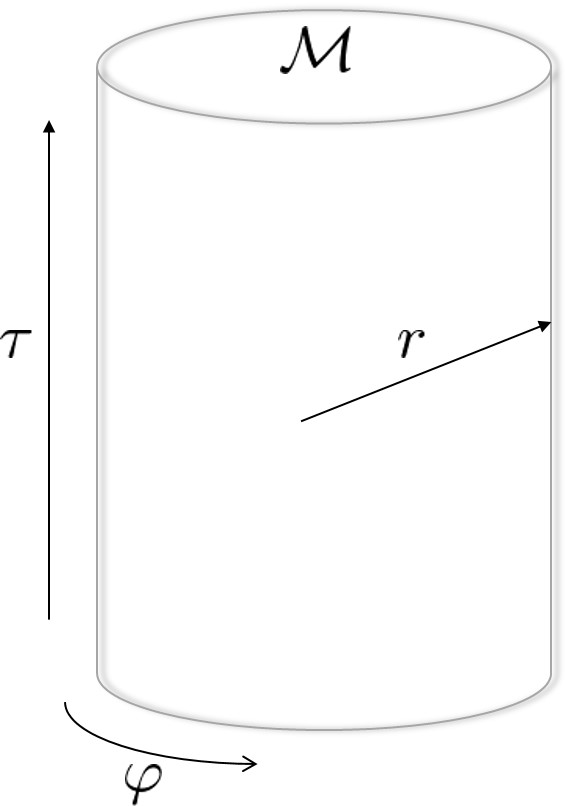}
\captionsetup{width=2\linewidth}
\caption{We consider the manifold $\rd M$ having the topology of a solid cylinder. Its boundary is taken to be the timelike cylinder at spatial infinity. }
\label{manifold}
\end{SCfigure}

Our three-dimensional manifold $\rd M$ is taken to have the topology of a cylinder $\mathbb{R} \times D_2$, where $\mathbb{R}$ is parametrized by the time coordinate $x^0\equiv\tau \equiv t/l$ and $D_2$ is the two-dimensional spatial manifold parametrized by coordinates $r$, $x^1\equiv \varphi$, with periodicity $\varphi \sim \varphi +2\pi$. We introduce light-cone coordinates $x^\pm\equiv \tau  \pm \varphi$ with $\partial_\pm=\frac{1}{2}(\partial_\tau\pm\partial_\varphi)$. The boundary $\partial \rd M$ of the spacetime at spatial infinity ($r=\infty$) is thus a timelike cylinder of coordinates $t,\varphi$, see Fig. \ref{manifold}. 

\subsection{Boundary conditions and phase space}
We adopt Fefferman-Graham coordinate system where the metric is given by ($i=0,1$) 
\be
ds^2=\frac{\ell^2}{r^2}dr^2+\gamma_{ij}(r,x^k)dx^i dx^j,
\label{FG}
\en
with the expansion, close to the the boundary $r \gt \infty$, $\gamma_{ij}=r^2 g_{ij}^{(0)}(x^k)+\rd O(1)$.
We call asymptotically AdS$_3$ spaces, in the sense of Brown-Henneaux \cite{BrownHenneaux}, metrics of the form \eqref{FG}, where the boundary metric $g_{ij}^{(0)}$ is fixed as
\be
g_{ij}^{(0)}dx^i dx^j=-dx^+ dx^-.
\label{BHbc}
\en
These \emph{Brown-Henneaux boundary conditions} are Dirichlet boundary conditions with a flat boundary metric \eqref{BHbc} on the cylinder located at spacial infinity. \\

It was shown in \cite{Banados4} that the most general solution (up to trivial diffeomorphisms) to Einstein's equations with $\Lambda=-1/\ell^2$ with boundary conditions \eqref{FG}, \eqref{BHbc} is 
\be
ds^2=\frac{\ell^2}{r^2}dr^2-\left(rdx^+ -\frac{\ell^2}{r}L(x^-)dx^-\right)\left(rdx^- -\frac{\ell^2}{r}\bar L(x^+)dx^+\right),
\en
where $L(x^-)$ and $\bar L (x^+)$ are two single-valued arbitrary functions of $x^-$ and $x^+$, respectively. In this gauge, one recovers well known geometries when these functions are constant; AdS$_3$ in global coordinates is recovered when $L=\bar L=-1/4$, $L=\bar L=0$ corresponds to the massless BTZ, while generic positive values of $L, \bar L$ correspond to generic BTZ geometries of mass $M=(L + \bar L)/(4G)$ and angular momentum $J=\ell(L-\bar L)/(4G)$.\\

Let us now translate these boundary conditions in the Chern-Simons formalism. We choose a dreibein $e^a$ which satisfies $ds^2=\eta_{ab}e^a e^b$ with an off-diagonal metric $\eta_{ab}$; see Appendix \ref{conventions} for our conventions. One can check that 
\be
e^0=-\frac{r}{\sqrt 2}dx^-+\frac{\ell^2}{\sqrt 2 r}\bar L(x^+)dx^+ \virg e^1=\frac{r}{\sqrt 2}dx^+ -\frac{\ell^2}{\sqrt 2 r}L(x^-)dx^-\virg e^2=\frac{\ell}{r}dr,
\en
reproduces $ds^2=2e^0e^1+(e^2)^2$, as desired.
Then, the torsion free first Cartan structure equation \eqref{torsion} determines uniquely the associated spin connections
\be
\om^0=\frac{r}{\sqrt 2 \ell}dx^-+\frac{\ell}{\sqrt 2 r}\bar L(x^+)dx^+ \virg \om^1=\frac{r}{\sqrt 2 \ell}dx^+ +\frac{\ell}{\sqrt 2 r}L(x^-)dx^- \virg \om^2=0.
\en

The corresponding chiral Chern-Simons flat\footnote{The fact that $A, \bar A$ are flat (i.e. $F=\bar{F}=0$) is ensured by the fact that $L(x^-), \bar L(x^+)$ are chiral.} connections $A=(\om^a+e^a/\ell)j_a$, $\bar A=(\om^a-e^a/\ell)j_a$ are therefore (with $j_a$ our generators)
\be
A=\left(
\begin{array}
[c]{cc}%
\dfrac{dr}{2r} & \dfrac{\ell}{r}\bar L(x^+)dx^+\\
\dfrac{r}{\ell}dx^+ & -\dfrac{dr}{2r}
\end{array}
\right)  \virg \bar A=\left(
\begin{array}
[c]{cc}%
-\dfrac{dr}{2r} & \dfrac{r}{\ell}dx^-\\
\dfrac{\ell}{r}L(x^-)dx^- & \dfrac{dr}{2r}
\end{array}
\right).
\label{onshellA}
\en

A very useful trick is to notice that one can factorize out the $r$-dependance of the gauge fields by performing the following gauge transformation:
\be
a=b^{-1}A \,b+b^{-1}db \virg \bar a=b\,\bar A \,b^{-1}+b\,db^{-1},
\label{aa}
\en
with 
\begin{equation}
b(r)=\left(
\begin{array}
[c]{cc}%
r^{-1/2} & 0\\
0 & r^{1/2}
\end{array}
\right).
\end{equation}
Indeed, one can check that the reduced connections $a, \bar a$ are $r$-independent; 
\[
a=\left(
\begin{array}
[c]{cc}%
0 & \ell \,\bar L(x^+)dx^+\\
dx^+/\ell & 0
\end{array}
\right)  \virg \bar a=\left(
\begin{array}
[c]{cc}%
0 & dx^-/\ell\\
\ell \, L(x^-)dx^- & 0
\end{array}
\right).
\label{onshella}
\]

In analogy with the on-shell reduced connections \eqref{aa}, we define the off-shell reduced gauge connections $a=a_\mu^a j_a dx^\mu$ and $\bar a=\bar a_\mu^a j_a dx^\mu$ as
\be
a=b^{-1}A \,b+b^{-1}db \virg \bar a=\bar b^{-1}\,\bar A \,\bar b+\bar b^{-1}\,d\bar b,
\en
such that $a_r=0=\bar a_r$. We impose our boundary conditions in the following way; they come in two sets:
\be
\badat{3}
&(i) \ \ \, a_-=0=\bar a_+ ,\\
&(ii)\ \ \, a_+=\frac{\sqrt 2}{\ell}j_1+0\,j_2+\sqrt 2 \ell L(x^+)j_0 , \ \ \ \bar a_-=\sqrt 2 \ell \bar L(x^-)j_1+0\,j_2+\frac{\sqrt 2}{\ell}j_0.
\label{bc}
\eadat
\en

The phase space is clearly contained in these boundary conditions, with $\bar b=b^{-1}$.
We will see that the first set of boundary conditions $(i)$ will reduce the Chern-Simons action to a sum of chiral $SL(2,\mathbb{R})$ Wess-Zumino-Witten (WZW) actions. The remaining set $(ii)$ will be used to further reduce the WZW model to Liouville theory. 

\subsection{Asymptotic symmetry algebra}
The asymptotic symmetries correspond to the set of gauge transformations\footnote{The analysis of asymptotic boundary conditions can also be performed from the geometrical point of view, in terms of the metric (or the vielbein and the spin connection). In that case, instead of studying the gauge transformations one studies the asymptotic Killing vectors that preserve the form of the metric (\ref{BHbc}) at large $r$.}
\be
\delta a=d\lambda+[a,\lambda] \virg \delta \bar a=d\bar \lambda+[\bar a,\bar \lambda] 
\en 
that preserve the asymptotic behavior of the connections $a$, $\bar a$, namely equations \eqref{bc}. Writing the gauge parameters $\lambda=\lambda^a j_a$, $\bar \lambda=\bar \lambda^a j_a$ we find\footnote{We compute for instance $\delta a_+=\partial_+ \lambda +[a_+,\lambda]$ and identify the components of left and right hand sides along the generators $j_a$.} that the latter have to be of the form 
\be
\badat{3}
&\lambda=\ell^2\left(L \lambda^1 - \frac{1}{2}\p_+^2 \lambda^1 \right)j_0 + \lambda^1 j_1 - \frac{\ell}{\sqrt 2}\p_+ \lambda^1 j_2, \\
&\bar \lambda=\bar \lambda^0 j_0 + \ell^2\left(\bar L \bar \lambda^0 - \frac{1}{2}\p_-^2 \bar \lambda^0 \right) j_1 + \frac{\ell}{\sqrt 2}\p_- \bar \lambda^0 j_2,
\label{lambda}
\eadat
\en
where the arbitrary functions $\lambda^1$, $\bar \lambda^0$ depend only on $x^+$, $x^-$ respectively\footnote{This follows directly from $a_-=0=\bar a_+$.}. Writing $Y \equiv \ell \lambda^1/\sqrt 2$, $\bar Y \equiv \ell \bar \lambda^0/\sqrt 2$, we find 
\be
\badat{3}
&\delta L=Y \p_+ L +2 L \, \p_+ Y-\frac{1}{2}\p_+^3 Y, \\
&\delta \bar L=\bar Y \p_- \bar L +2 \bar L \, \p_- \bar Y-\frac{1}{2}\p_-^3 \bar Y.
\label{deltaL}
\eadat
\en
At this stage, we can already notice that $L$ and $\bar L$ transform in the same way as a two-dimensional CFT energy-momentum tensor does under generic infinitesimal conformal transformations, and one can already see that the last term, associated to the Schwarzian derivative, indicates the presence of a central extension.

The variation of the canonical generators associated to the asymptotic symmetries spanned by $\lambda$ take a very simple form in the Chern-Simons formalism \cite{Banados2,Troncoso1}; one has
\be
\delta Q[\lambda]= -\frac{k}{2 \pi} \int_0^{2\pi} (\lambda, \delta a_+) \,d\varphi \virg \delta \bar Q[\bar \lambda]= -\frac{k}{2 \pi} \int_0^{2\pi} (\bar \lambda, \delta a_-) \,d\varphi.
\label{deltaQ}
\en
One then find that expressions in \eqref{deltaQ} become linear in the deviation of the fields, so that they can be directly integrated as 
\be
Q_Y= -\frac{k}{2 \pi} \int_0^{2\pi} Y \, L \,d\varphi , \ \ \ 
\bar Q_{\bar Y}= -\frac{k}{2 \pi} \int_0^{2\pi} \bar Y \, \bar L \,d\varphi.
\en

The Poisson brackets fulfill $\delta_{Y_1}Q_{Y_2}=\{Q_{Y_2},Q_{Y_1}\}$; therefore, the algebra of the canonical generators can be directly computed from the transformation laws \eqref{deltaL}. Defining the modes  ($m \in \mathbb Z$)
\be
L_m \equiv  \frac{k}{2 \pi} \int_0^{2\pi} e^{im \varphi} \, L \,d\varphi \pvirg \bar L_m \equiv \frac{k}{2 \pi} \int_0^{2\pi} e^{im \varphi} \, \bar L \,d\varphi,
\en
one finds 
\be
\badat{3}
&i\{L_m,L_n\}=(m-n)L_{m+n}+\frac{c}{12}m^3 \delta_{m+n,0}, \\
&i\{L_m,\bar L_n\}=0,\\
&i\{\bar L_m,\bar L_n\}=(m-n)\bar L_{m+n}+\frac{\bar c}{12}m^3 \delta_{m+n,0},
\label{Vir}
\eadat
\en
with central elements given by
\be 
c=\bar c =6k=\frac{3 \ell}{2G}.
\label{centralcharge}
\en

This shows that the charge algebra associated to the symmetry transformations that preserve the AdS$_3$ asymptotic form consists of a direct sum of two copies of the \emph{Virasoro algebra}, with $c$ being the \emph{central charge}. The Virasoro algebra \eqref{Vir} is the algebra of local conformal transformations in two-dimensions; it is the central extension of the Witt algebra and is \emph{infinite}-dimensional (recall that $m \in \mathbb Z$). Notice that the standard redefinitions $L_m \gt L_m+\frac{c}{24}$, $\bar L_m \gt \bar L_m+\frac{\bar c}{24}$ change the central terms in the algebra to $\frac{c}{12}m(m^2-1) \delta_{m+n,0}$ and $\frac{\bar c}{12}m(m^2-1) \delta_{m+n,0}$. 
From the CFT perspective, the existence of a non-trivial central extension is the result of a conformal (or Weyl) anomaly in the quantum theory. However, since this derivation was purely classical, finding a central extension at the classical level is quite remarkable. In the context of AdS/CFT, this is interpreted as a classical bulk (AdS$_3$) computation that describes a property of the effective action of the quantum boundary (CFT$_2$) theory. Algebra \eqref{Vir} was first shown in the seminal paper of Brown and Henneaux \cite{BrownHenneaux} in 1986, and this is the reason why this result is considered as the precursor of the AdS/CFT correspondence; notice that for this reason the central charge \eqref{centralcharge} is often called the Brown-Henneaux central charge.

It is worth mentioning that in the semi-classical limit $\ell \gg G$ (recall that the Planck length in three dimensions is $\ell_P \sim G$), the central charge \eqref{centralcharge} tends to infinity.
Also, notice that the quantization of the Chern-Simons level $k$ implies that the quantum theory seems to be well defined only for discrete values of the dimensionless ratio $\ell/G$ and, hence, discrete values of the central charge. This discretization of $c$ is also understood from the dual CFT point of view, since the Zamolodchikov $c$-theorem prohibits the central charge to be continuous \cite{ctheorem}. 

Recently, it was shown that the asymptotic symmetries of three-dimensional gravity with Brown-Henneaux boundary conditions can be defined everywhere into the bulk of spacetime \cite{sympl1}, promoting in this sense the two copies of Virasoro algebra \eqref{Vir} to a new kind of symmetries, the so-called symplectic symmetries \cite{sympl2}. These symplectic symmetries are large gauge transformations that are defined everywhere in spacetime, not only in an asymptotic region. This result suggests that the dual two-dimensional CFT is not necessarily only defined at the boundary, since the surface charges and algebra are defined on any circle located in the AdS bulk. The appearance of symplectic symmetries is related to the existence of a presympletic form which vanishes on-shell and is most likely conditioned by the absence of propagating degrees of freedom in the bulk.


\section{A brief introduction to Wess-Zumino-Witten models}
\label{secWZW}

In this section, we will present an important ingredient in our discussion, the Wess-Zumino-Witten model, which appears as an intermediate step in the connection between Chern-Simons and Liouville actions.

\subsection{The nonlinear sigma model} 

In quantum field theory, a \textit{nonlinear sigma model}\footnote{The name comes from the fact that this model appeared for the first time in the description of a spinless meson called $\sigma$.} describes scalar fields $\phi^i$ ($i=1,...,n$) as maps from a flat spacetime to a target manifold. The latter is a $n$-dimensional Riemannian manifold $\rd M_n$ equipped with a metric $g_{ij}(\phi)$ which depends on the fields, this is why the model is intrinsically nonlinear. In other words, the coordinates on $\rd M_n$ are the scalar fields $\phi(x)$, in which the $x^\mu$, $\mu=1,...D$ are the Cartesian coordinates of a flat spacetime. An action for this model is given by
\be
S_\sigma[\phi]=\frac{1}{4 a^2}\int d^Dx \,g_{ij}(\phi)\eta^{\mu \nu}\p_\mu \phi^i \p_\nu \phi^ i,
\en
with $a^2>0$ a dimensionless coupling constant.

The so-called Wess-Zumino-Witten model\footnote{It is important not to mistake the Wess-Zumino-Witten model with the Wess-Zumino model that describes four-dimensional supersymmetric interactions. The former is usually referred to as the Wess-Zumino-Novikov-Witten model.} involves a particular two-dimensional $\sigma $-model in which the role of the target space is played by a semi-simple Lie group $G$ and the fields are matrix fields living on $G$, noted $g(x)$. For a two-dimensional manifold $\Sigma$  with coordinates $x^0=\tau $, $x^1=\varphi$ ($\mu, \nu= 0,1$), the action of this nonlinear sigma model takes the form \cite{DiFrancesco}
\be
S_\sigma[g]=\frac{1}{4 a^2}\int_{\Sigma} d^2x \,\mathrm{Tr}\left[\eta^{\mu \nu}\p_\mu g \, \p_\nu (g^{-1}) \right].
\label{sigma}
\en
The group $G$ has to be semi-simple to ensure the existence of the trace Tr, but can be either compact or non-compact.

This theory is conformally invariant only at the classical level. Indeed, under quantization, the coupling $a$ acquires a scale dependance, leading therefore to a non-vanishing $\beta$-function (the quantum theory is in fact asymptotically free). Furthermore, even at the classical level, this theory is not totally satisfactory since it does not possess two conserved currents that factorize into a left (or holomorphic) and a right (antiholomorphic) part; this is the fundamental property of holomorphic factorization of a CFT. Indeed, the equations of motion are\footnote{Notice the useful relation $\delta (g^{-1})=-g^{-1}\delta g \, g^{-1}$, which can be derived from $\delta(g g^{-1})=0$.}
\be
\p^\nu(g^{-1}\p_\nu g)=0,
\en
which read in light-cone coordinates $x^\pm\equiv \tau \pm \varphi$, 
\be
\p_+ J_-+\p_-J_+=0,
\label{pp}
\en
where we have defined the currents as $J_+\equiv g^{-1}\p_+ g$, $J_-\equiv g^{-1}\p_- g$. We thus see that the equations of motion derived from \eqref{sigma} do not lead to the independent conservation of the left and right currents $J_\pm$. If one is conserved, \eqref{pp} implies that the other current has to be conserved as well.

\subsection{Adding the Wess-Zumino term} 

In order to have two independently conserved currents, it has been observed in \cite{WittenWZW,Novikov} that one has to consider, instead, the more involved action 
\be
S=S_\sigma[g]+k \Gamma[G],
\label{S1}
\en
with $k$ an integer\footnote{The name is of course not innocent, since we will see that it is indeed related to the Chern-Simons level $k$.}, and with the Wess-Zumino term $\Gamma[G]$ being
\be
\badat{2}
\Gamma[G]&=\frac{1}{3} \int_{V} d^3x \,\ep^{\mu \nu \rho}\mathrm{Tr}\left[G^{-1}\p_\mu G \,G^{-1}\p_\nu G \,G^{-1}\p_\rho G \right]\\
&\equiv \frac{1}{3} \int_{V} \mathrm{Tr}\left[(G^{-1}dG)^3 \right],
\label{G}
\eadat
\en
where $V$ is a three-dimensional manifold having $\Sigma$ as a boundary, $\p V= \Sigma$, and $G$ is the extension of the element $g$ on $V$. Notice that there are of course several choices for a $V$ extending $\Sigma$, leading therefore to a potential ambiguity in the definition of $\Gamma$. In fact, for the quantum theory to be well-defined, and depending on the Lie group considered, this can imply a quantization condition for $k$. However, in the case of $SL(2,\mathbb{R} )$ we are interested in, this issue does not appear.

Action \eqref{S1} may look surprising since it mixes a nonlinear sigma model in two dimensions with the three-dimensional action \eqref{G}. However, the Wess-Zumino term has the fundamental property that its variation under $g \gt g + \delta g$ yields a two-dimensional functional. Actually, one can show that its variation is a total derivative, leading to the result (using Stokes' theorem)
\be
\delta \Gamma[G]= \int_\Sigma d^2 x  \,\mathrm{Tr}\left[\ep^{\mu \nu}\delta g g^{-1}\p_\mu g \,g^{-1}\p_\nu g g^{-1}\right].
\en
Using this result, one can see that the equations of motion derived from \eqref{S1} read
\be
\frac{1}{2a^2}\eta^{\mu \nu}\p_\mu(g^{-1}\p_\nu g)-k \ep^{\mu \nu} \p_\mu(g^{-1}\p_\nu g)=0.
\label{eom}
\en
In light-cone coordinates $x^\pm$ (see Appendix \ref{conventions} for our conventions), they become
\be
(1-2a^2k)\p_+(g^{-1}\p_- g)+(1+2a^2k)\p_-(g^{-1}\p_+ g)=0.
\en
Therefore, for $a^2=-1/(2k)$, which implies $k<0$, one finds the conservation of the current $\p_+J_-=0$, while for $a^2=1/(2k)$, one finds the conservation of the dual current $\p_- J_+=0$. For the same conditions, one can show that the beta-function vanishes \cite{WittenWZW}, representing a conformal invariant fixed point.

Taking $a^2=-1/(2k)$, one obtains the \textit{Wess-Zumino-Witten} (WZW), or \textit{Wess-Zumino-Novikov-Witten} (WZNW) action; namely
\be
S_{\text{WZW}}[g]=\frac{k}{2}\int d^2x \,\mathrm{Tr}\left[\eta^{\mu \nu}g^{-1}\p_\mu g \,g^{-1}\p_\nu g \right]+k\Gamma[G].
\label{WZW}
\en

This action is sometimes called \textit{non-chiral} WZW action, since it does not distinguish between left and right movers (it is symmetric under $x^+ \leftrightarrow x^-$), unlike the chiral action we will present later on. The solution of the equations of motion derived from \eqref{WZW}, namely $\p_+(g^{-1}\p_- g)=0$, is simply
\be
g=\theta_+(x^+)\theta_-(x^-),
\label{movers}
\en
where $\theta_+(x^+)$ and $\theta_-(x^-)$ are arbitrary functions. Equation \eqref{movers} means that left and right movers do not interfere between each others. One checks that the model described by \eqref{WZW} has the two conserved currents
\be
J_- \equiv g^{-1}\p_- g \virg \bar J_+ \equiv -\p_+ g \, g^{-1}.
\en
The independent conservation of the two currents implies that action \eqref{WZW} is invariant under $g \gt \Theta_+(x^+) g \Theta_-^{-1}(x^-)$, with $\Theta_\pm$ two arbitrary matrices valued in $G$. Therefore, one sees that the global $G \times G$ invariance of the sigma model has been promoted to a local $G(x^+) \times G(x^-)$ invariance.

\section{From Chern-Simons to Wess-Zumino-Witten}
\label{secCSWZW}
Chern-Simons theories have been shown to reduce to Wess-Zumino-Witten theories on the boundary \cite{Seiberg1, Seiberg2}. In particular, we will see explicitly in this section that the Chern-Simons theory 
\be
S_E[A,\bar A]=S_{CS}[A]-S_{CS}[\bar A],
\label{SE}
\en
with\footnote{Notice that we have changed the overall sign for latter convenience.}
\be
S_{CS}[A]=-\frac{k}{4\pi}\int_{\rd M} \mathrm{Tr} \left[A \wedge dA +\frac{2}{3}A\wedge A\wedge A \right],
\label{CS}
\en
describing $(2 + 1)$-dimensional gravity with $\Lambda=-1/\ell^2$ reduces under our boundary conditions to the $SL(2,\mathbb{R})$ WZW model on the cylinder at spatial infinity.
From now on, we will use, instead of the level $k=\ell/(4 G)$, the constant
\be
 \kappa \equiv \frac{k}{4\pi}= \frac{\ell}{16 \pi G}.
\label{k}
\en

Again, let us emphasize the advantage of the Chern-Simons formulation: Instead of working with a second order action in terms of the metric, we work with two flat gauge connections $A, \bar A$. In this section, we want to show explicitly how the first set of boundary conditions \eqref{bc} implements the reduction of the Chern-Simons action \eqref{SE} to a sum of two chiral Wess-Zumino-Witten actions.

\subsection{Improved action principle}
At this stage, we have a small issue to solve: our boundary conditions \eqref{bc} do not lead to a well-defined action principle. To see that, let us first rewrite our Chern-Simons action \eqref{CS} explicitly in coordinates $r, \tau, \varphi$:
\begin{multline}
S_{CS}[A]
=-\kappa \int_{\rd M} dx^\mu \wedge dx^\nu \wedge dx^\rho \, \mathrm{Tr} \big [A_\mu \partial_\nu A_\rho +\frac{1}{3} A_\mu [A_\nu, A_\rho] \big]\\
=-\kappa \int_{\rd M} dr d\tau  d\varphi  \,\mathrm{Tr} \big[A_r (\partial_\tau A_\varphi - \partial_\varphi A_\tau)+A_\tau (\partial_\varphi A_r -\partial_r A_\varphi)+  A_\varphi (\partial_r A_\tau - \partial_\tau A_r)\\
+2A_\tau[A_\varphi,A_r] \big].
\end{multline}
Integrating by parts the second and fifth terms, and keeping only the radial boundary terms (the ones on $\varphi$ vanish because of periodicity), one has, using Stokes' theorem,
\begin{multline}
S_{CS}[A]=-\kappa \int_{\rd M} dr  d\tau  d\varphi  \,\mathrm{Tr} \big[A_r \dot A_\varphi - A_\varphi \dot A_r+2 A_\tau (\partial_\varphi A_r -\partial_r A_\varphi+ [A_\varphi,A_r])\big]\\
+\kappa \int_{\partial \rd M} d\tau  d\varphi  \,\mathrm{Tr} \big[A_\tau A_\varphi \big],
\end{multline}
where the dot stands for $\p_\tau$. The last boundary contribution being irrelevant, one can reabsorb it in the definition of the action. Therefore, the final form for the Chern-Simons action in terms of coordinates is given by
\be
S_{CS}[A]=-\kappa \int_{\rd M} dr  d\tau  d\varphi \, \mathrm{Tr}  \left[A_r \dot A_\varphi-A_\varphi \dot A_r+2A_\tau F_{\varphi r}\right],
\label{CScoord}
\en
where $F=dA+A\wedge A$ is the curvature two-form associated to the connection. However, this action does not have a well-defined action principle. Indeed, computing the variation of the total action \eqref{SE}, one finds
\be
\delta S_E=(\text{EOM}) + 2 \kappa \int_{ \partial \rd M} d\tau  d\varphi \mathrm{Tr} \left[A_\tau \delta A_\varphi - \bar A_\tau \delta \bar A_\varphi  \right],
\en
which is not zero on-shell when $A_-$ and $\bar A_+$ are required to vanish on the boundary. Therefore, in order to have a well-defined variational principle, one must add the following surface term to the action
\be
I= - \kappa \int_{ \partial \rd M} d\tau d\varphi \mathrm{Tr} \left[A_\varphi^2 +\bar A_\varphi^2  \right],
\en
which is such that the improved action 
\be
S[A, \bar A]\equiv S_E+I = S_{CS}[A]-S_{CS}[\bar A] - \kappa \int_{\partial \rd  M} d\tau  d\varphi \mathrm{Tr} \left[A_\varphi^2 +\bar A_\varphi^2  \right]
\label{SAAb}
\en
satisfies $\delta S[A,\bar A]=0$ (recall that $A_-=0=\bar A_+$ on the boundary imply $A_\tau=A_\varphi$, $\bar A_\tau=\bar A_\varphi$).

\subsection{Reduction of the action to a sum of two chiral WZW actions}
Now that we have an action with a well-defined variational principle, we are ready to reduce the Chern-Simons to the WZW model. First, we will focus on the chiral sector, and then we will do a similar computation for the anti-chiral sector to finally compose the full WZW action.\\

The chiral part of the improved action \eqref{SAAb} is given by
\be
S[A]\equiv S_{CS}[A]- \kappa \int_{\partial \rd M} d\tau  d\varphi \mathrm{Tr} \left[A_\varphi^2\right],
\label{chiral}
\en
with $S_{CS}[A]$ given by \eqref{CScoord}. Looking at the last term of \eqref{CScoord}, we realize that the component $A_\tau$ of the connection merely plays the role of a Lagrange multiplier, implementing the constraint $F_{r\varphi}=0$. Therefore, assuming no holonomies (i.e. no holes in the spatial section), one can solve this constraint as
\be
A_i=G^{-1}\partial_i G \virg (i=r,\varphi),
\label{eq1}
\en
where $G$ is an $SL(2,\mathbb R)$ group element. Indeed, the solution to $dA+A\wedge A=0$ is locally\footnote{Since we do not consider holonomies, we assume that this will also hold globally. In general, one should consider the more general case with holonomies, which appear as additional zero-modes that one should take into account if one wants to describe three-dimensional black holes. A treatment of holonomies was considered in \cite{Maoz, Spindel}.} given by $A=G^{-1}dG$. The condition $F_{r\varphi}= 0$, as a first class constraint, generates gauge transformations. One can partially fix the gauge by imposing\footnote{In fact, this consists of a gauge fixing condition only in an off-shell formulation, since this relation is obviously satisfied for the on-shell connection \eqref{onshellA}.} $\partial_\varphi A_r=0$, which allows to factorize the general solution into
\be
G(\tau,r,\varphi)= g(\tau,\varphi)\, h(r,\tau).
\label{eq2a}
\en
This implies $A_r=h^{-1}\partial_r h$ and $A_\varphi=h^{-1}g^{-1}g^\prime h$, with the prime standing for the derivative with respect to $\varphi$. We will assume that, as it happens for the solutions of interest, $\dot h|_{\partial \rd M}=0$.
We are now ready to reduce the chiral action \eqref{chiral}. Plugging \eqref{eq1} and \eqref{eq2a}, the boundary term simply reads
\be
- \kappa \int_{\partial \rd M} d\tau  d\varphi \mathrm{Tr} \left[(g^{-1}\partial_\varphi g)^2\right],
\label{eq2b}
\en
while the three-dimensional term gives explicitly (using the constraint $F_{r\varphi}=0$)
\be
\badat{2}
S_{CS}[A]&=\kappa \int_{\rd M} dr d\tau d\varphi \, \mathrm{Tr}  [ \partial_r h h^{-1} \dot h h^{-1} g^{-1} g'+\partial_r h h^{-1} g^{-1} \dot g g^{-1} g'\\
& -\partial_r h h^{-1} g^{-1} \dot g' -h^{-1} \partial_r h h ^{-1} g^{-1} g' \dot h - h^{-1} g^{-1} g' \dot h h^{-1} \partial_r h + h^{-1} g^{-1} g' \partial_r \dot h].
\label{eq3}
\eadat
\en
On the other hand, one can see that, with the convention $\ep^{r t \varphi}\equiv 1$, 
\be
\badat{2}
\frac{1}{3} \kappa \int_{\rd M} \mathrm{Tr} \left[(G^{-1} dG )^3\right]=\kappa \int_{\rd M} dr  d\tau  d\varphi \, \mathrm{Tr}&[\partial_r h h^{-1} \dot h h^{-1} g^{-1} g'+\partial_r h h^{-1} g^{-1} \dot g g^{-1} g' \\
&- \partial_r h h^{-1} g^{-1}g' g^{-1} \dot g -h^{-1} \partial_r h h ^{-1} g^{-1} g' \dot h ].
\eadat
\en
Integrating by parts the third term in \eqref{eq3}, one finds
\be
\badat{2}
S_{CS}[A]=\frac{\kappa}{3} \int_{\rd M} \mathrm{Tr} \left[(G^{-1} dG )^3\right]+\kappa \int_{\rd M} dr  d\tau  d\varphi \, \mathrm{Tr} & [- h^{-1} g^{-1} g' \dot h h^{-1} \partial_r h + h^{-1} g^{-1} g' \partial_r \dot h ].
\eadat
\en
Finally, realizing that the last two terms are nothing but
\be
\partial_r(G^{-1}\partial_\varphi G  G^{-1}\partial_\tau G)=\partial_r(h^{-1}g^{-1} g' h(h^{-1} g^{-1} \dot g h+h^{-1} \dot h)),
\en
and recalling that $\dot h=0$ on $\partial \rd M$, we have (using Stokes' theorem again) 
\be
\badat{2}
S_{CS}[A]&=\kappa \int_{\partial \rd M} d\tau d\varphi \, \mathrm{Tr}  \left[g^{-1} \partial_\varphi g g^{-1} \partial_t g\right]+\frac{\kappa}{3} \int_{\rd M} \mathrm{Tr} \left[(G^{-1} dG )^3\right].
\eadat
\en

Therefore, we have shown that the Chern-Simons action for the chiral copy \eqref{chiral} reduces to
\be
\badat{2}
S[A]&=\kappa \int_{\partial \rd M} d\tau d\varphi \, \mathrm{Tr}  \left[g^{-1} \partial_\varphi g (g^{-1} \partial_t g- g^{-1} \partial_\varphi g)\right]+\kappa \Gamma[G]. \\
&\equiv S^R_\mathrm{WZW}[ g],
\eadat
\label{cWZW}
\en
This is a \textit{chiral} Wess-Zumino action action for the group element $g$. In light-cone coordinates $x^\pm = \tau  \pm \varphi$, \eqref{cWZW} reads 
\be
S[A]=2 \kappa \int_{\partial \rd M} d\tau  d\varphi \, \mathrm{Tr}  \left[g^{-1} \partial_\varphi g \,g^{-1} \partial_- g\right]+\kappa \Gamma[G].
\en
This first order action describes a \textit{right-moving} group element $g$; this is the reason for the name ``chiral'', which means that the action distinguishes between left and right movers. Indeed, the equations of motion are $\partial_-(g^{-1}g')=0$, whose solution is given by $g=f(\tau)k(x^+)$, which is the equation for an element moving along the $x^+$ direction\footnote{A right mover is often denoted $g(x^+)$; it is an abuse of notation to mean that $g$ moves along the $x^+$ direction.}. \\

Similarly, the anti-chiral action 
\be
S[\bar A] \equiv S_{CS}[\bar A] + \kappa \int_{\partial \rd  M} d\tau d\varphi \mathrm{Tr} \left[\bar A_\varphi^2 \right],
\en
after solving $\bar F_{r \phi}= 0$ by $\bar A_i=\bar G^{-1} \partial_i \bar G$, $\bar G(t,r,\varphi)= \bar g(t,\varphi)\, \bar h(r,t)$ (with $\dot{\bar h}=0$ on $\p \rd M$), can be written as
\be
S[\bar A]=-\kappa \int_{\rd M} dr  d\tau d\varphi \, \mathrm{Tr}  \left[\bar A_r \dot {\bar {A_\varphi}}-\bar A_\varphi \dot {\bar{A_r}}\right]+ \kappa \int_{\rd \partial \rd M} d\tau  d\varphi \mathrm{Tr} \left[\bar A_\varphi^2 \right].
\en

The only difference with the chiral action being the sign of the two-dimensional term, one finds easily that
\be
\badat{2}
S[\bar A]&=\kappa \int_{\partial \rd M} d\tau d\varphi \, \mathrm{Tr}  \left[\bar g^{-1} \partial_\varphi \bar g (\bar g^{-1} \partial_t \bar g+ \bar g^{-1} \partial_\varphi \bar g)\right]+\kappa \Gamma[\bar G]\\
&\equiv S^L_{\mathrm{WZW}}[\bar g],
\eadat
\en
where $S^L_{WZW}[\bar g]$ denotes a WZW action for a left-moving element $\bar g$. Indeed, the equations of motion $\p_+(\bar g^{-1}\bar g')=0$ imply $\bar g=\bar f(\tau)\bar k(x^-)$.\\

Therefore, combining left and right sectors, we have shown that the total Chern-Simons action is given by
\be
S[A, \bar A]=S^R_\mathrm{WZW}[g]-S^L_\mathrm{WZW}[\bar g].
\label{SRL}
\en

\subsection{Combining the sectors to a non-chiral WZW action}
In order to recover the standard (non-chiral) WZW action \eqref{WZW}, one can use the Hamiltonian form, since the chiral and anti-chiral actions are linear and of first order in time derivative.
We combine left and right movers as $k \equiv g^{-1} \bar g$, $K=G^{-1} \bar G$; we define as well
\be
\Pi \equiv -\bar g^{-1}\p_\varphi g g^{-1}\bar g - \bar g^{-1}\p_\varphi \bar g,
\en
and observe that
\be
\Gamma[K]  = -\Gamma[G] + \Gamma[\bar G]- \int_{\partial \rd M} \text{Tr} \left(d\bar g \bar g^{-1} d g g^{-1}\right).
\en
We are allowed to change the variables from $g$ and $\bar g$ to $k $ and $\Pi$. In terms of the latter, the action \eqref{SRL} reads
\be
S[k,\Pi] = \kappa \int_{\partial \rd M} d\tau  d\varphi\,  \text{Tr}\, \left[ \Pi k^{-1} \dot k-\frac{1}{2} (\Pi^2 + ( k^{-1} k')^2) \right] \, - \frac{\kappa}{3} \int_{\rd M}  \text{Tr} \,\left[(K^{-1}dK)^3 \right] .
\en
Eliminating the auxiliary variable $\Pi$ by using its equation of motion, one finally gets
\be
S[k] = \kappa  \int_{\partial \rd M} d\tau d\varphi\,  \text{Tr} \, \left[2 k^{-1}\p_+ k k^{-1} \p_- k \right] - \frac{\kappa }{3} \int_{\rd M} \text{Tr}\,  \left[(K^{-1}dK)^3 \right],
\label{2der}
\en
which is the standard non-chiral $SL(2,\mathbb R)$ WZW action for an element $k$.

Notice that the above change of variables above is not well-defined for the zero modes \cite{Hernan3}. As a consequence, the equivalence of the sum of two chiral models with the non-chiral theory is not valid in that sector.


\section{From the WZW model to Liouville theory}
\label{secWZWLiouv}
So far, we have shown that the asymptotic dynamics of three-dimensional gravity with $\Lambda <0$ is described by the (non-chiral) WZW action for $SL(2,\mathbb R)$. However, only the first set of boundary conditions \eqref{bc} was used so far. In this section, we will see how the use of the second set further reduces the WZW model to eventually get Liouville field theory. Liouville theory is a two-dimensional conformal invariant field theory whose origin can be traced back to the work of Joseph Liouville in the 19th century. We will give a brief introduction to the classical and quantum Liouville theories in the last section.

\subsection{The Gauss decomposition}
In order to perform the reduction at the level of the action, it is useful to express the WZW action \eqref{2der} in local form upon performing a Gauss decomposition of the form
\be
\badat{2}
\label{gauss2}
K&=e^{\sqrt 2 X j_0}e^{\phi j_2}e^{\sqrt 2 Y j_1}\\
&= \left(  \begin{array}{cc} 1 & X \\ 0 & 1 \end{array} \right) \left(  \begin{array}{cc} e^{\frac{1}{2} \phi} & 0 \\ 0 & e^{-\frac{1}{2} \phi} \end{array} \right) \left(  \begin{array}{cc} 1 & 0 \\ Y & 1 \end{array} \right) \, ,
\eadat
\en
where $ X, Y, \phi$ are fields that depend on $u,\varphi$ and $r$. We assume that the decomposition holds globally (for subtleties in the presence of global obstructions, see \cite{Balog1}). The Gauss decomposition allows to rewrite the three-dimensional integral in \eqref{2der} as a two-dimensional integral using the relation 
\be
-\frac{1}{3}\text{Tr} ( K^{-1} dK)^3 = dr d\tau d\varphi \ \ep^{\alpha\beta\gamma}\, \p_\alpha \, \left( e^{-\phi} \p_\beta X\,  \p_\gamma Y \right)\, .
\en
Therefore, one finds (keeping only the radial boundary term)
\be
-\frac{1}{3} \int_{\rd M} \text{Tr} ( K^{-1} dK)^3 = \int_{\p \rd M} d\tau \, d\varphi\, 2 e^{-\phi} (\p_-X \p_+Y -\p_+ X \p_-Y).
\en
The two-dimensional integral in \eqref{2der} can be rewritten equivalently by replacing $k$ by $K|_{\p \rd M}$ since all factors of $h,\bar h$ exactly cancel in the trace. Therefore, one finds
\be
\int_{\partial \rd M} d\tau \, d\varphi\, \text{Tr} \, \left[2 k^{-1}\p_+ k k^{-1} \p_- k \right]=\int_{\partial \rd M} d\tau \,d\varphi\, \left(\p_-\phi \p_+ \phi +2 e^{-\phi}(\p_+X\p_-Y+\p_-X \p_+ Y)\right).
\en
One can then combine all terms and find that \eqref{2der} reduces to
\be
\label{totalhat}
S_{\mathrm{red}}  = 2 \kappa \int_{\p \rd M} d\tau \, d\varphi\,  \left(\frac{1}{2} \p_- \phi \, \p_+ \phi + 2 e^{-\phi}  \p_- X \, \p_+ Y  \right) \, ,
\label{reduced}
\en
where all fields $X, Y, \phi$ have  been pull-backed on $\p \rd M$.

\subsection{Hamiltonian reduction to the Liouville theory}

The second set of boundary conditions \eqref{bc} on the gauge fields set the currents of the WZW model to constants. This is the well-known Hamiltonian reduction of the WZW model to Liouville \cite{DS, Balog2, AS}.

Let us begin by considering the left and right moving WZW currents. They are given by\footnote{In this section, all the $k$ symbols appearing denote the group element $k = g^{-1}\bar g$, which has not to be confused with the level, which is a constant labeled by the same letter.}
 \begin{align}
J_a=k^{-1}\partial_a k ,
\qquad \bar J_a=-\partial_a k k^{-1}. 
\end{align}
Using the definition of $k$, we deduce (recall that $a=g^{-1}dg, \bar a=\bar g^{-1}d\bar g$):
\begin{align}
J_- &=
 -k^{-1} a_- k + \bar a_-  ,\qquad 
\bar J_+ =  a_+ - k \bar a_+ k^{-1} .
\end{align}
Then, using the first set of boundary conditions $(i)$, $a_-=\bar a_+=0$, we obtain a simple relation between $k$, the WZW currents, and the gauge fields: $J_- = k^{-1}\partial_- k=\bar a_-$, $\bar J_+ = -\partial_+ k k^{-1}= a_+$. Implementing the second set of boundary conditions $(ii)$, $a_+=({\sqrt 2}/{\ell})j_1+0\,j_2+\sqrt 2 \ell L(x^+)j_0$, $\bar a_-=\sqrt 2 \ell \bar L(x^-)j_1+0\,j_2+({\sqrt 2}/{\ell})j_0$, one finds
\be
\badat{2}
& J_-^0 =[k^{-1}\p_- k]^0 =\frac{\sqrt 2}{\ell} \virg J_-^2=0 ,\\
& \bar J_+^1 =[-\p_+ k k^{-1}]^1 =\frac{\sqrt 2}{\ell} \virg\bar J_+^2=0 .
\label{cons}
\eadat
\en
We thus see that the second set of boundary conditions \eqref{bc} has for effect to set the WZW currents to constants.
The set of constraints \eqref{cons} is equivalent, in terms of the $\phi,X,Y$ fields introduced above, to the set
\be
\badat{2}
& e^{-\phi} \p_- X = \frac{1}{\ell} \virg e^{-\phi}\p_+ Y = -\frac{1}{\ell}, \\
& X= 2 \ell \partial_+ \phi \virg Y= -2 \ell\, \partial_- \phi .
\eadat
\en

Before solving these constraints in the action, one has to be sure that the latter has a well-defined variational principle. To achieve so, one needs to add an improvement term to the action \eqref{reduced} as follows\footnote{This boundary term comes from the fact that the constraints restrict $\p_+ Y$ and $\p_- X$ rather than $X$ and $Y$ themselves.}
\begin{equation}\label{improved}
S_{\rm{impr}}\equiv S_{\mathrm{red}}-2 \kappa \int_0^{2\pi} d\varphi\ \Big(e^{-\phi} \left( X\partial_+ Y+Y\partial_-X\right)\Big)\Big|_{\tau_1}^{\tau_2}\, .
\end{equation}
After inserting the constraints, we are left with the Liouville action
\begin{equation}
S_{\rm{Liouville}}[\phi]=2 \kappa \int_{\p \rd M} d\tau\, d\varphi\ \left(\frac{1}{2}\partial_+\phi\, \partial_-\phi + \frac{2}{\ell^2}\exp({\phi})\right)\ .
\label{final}
\end{equation}
Notice that the boundary term in \eqref{improved} contributes as $2 \kappa \int_{\p \rd M} d\tau d\varphi \, ({4}/{\ell^2}) \exp({\phi})$. Also, notice that by shifting $\phi$ by a constant, one can set $2/\ell^2$ to any (positive) value.\\

We have therefore shown that the boundary dynamics of AdS gravity in $D=3$ dimensions is described by the two-dimensional Liouville action. Liouville theory is a conformal field theory and, therefore, has associated two mutually commuting sets of Virasoro generators $L_m$ and $\bar L_m$. Then, identifying these generators with the ones appearing in the asymptotic analysis carried out in Section \ref{secBC} is very tempting. Does it mean that Liouville theory \textit{is} the dual conformal theory of three-dimensional gravity with $\Lambda <0$? Actually not. The reduction we have presented is a classical computation, working at the level of the actions through the prescription of specific boundary conditions; to establish a correspondence between the two theories one would also need a full understanding at the quantum level. Nevertheless, the connection between Einstein theory on AdS$_3$ and Liouville theory on the boundary we have described, shows that the latter theory is, at least, an effective theory of the holographic dual CFT$_2$. But, before anticipating this discussion, let us briefly introduce the Liouville field theory for the readers who are not familiar with it.

\section{Liouville field theory}
\label{secLiouv}

\subsection{At classical level}
The Liouville differential equation was introduced in the 19th century in the context of the uniformization theorem for Riemann surfaces \cite{Lut,Picard1,Picard2}. This is a classical problem of mathematics that can be rephrased as the following question: In a two-dimensional space equipped with a metric $g_{\mu \nu}$, does it exist a function $\Omega $ such that a new metric $\tilde g_{\mu \nu} \equiv \Omega g_{\mu \nu}$ has a constant scalar curvature $\tilde R$? The answer turns out to be yes. To see this, one first defines $\Omega=e^{2 \phi}$ and, then, finds that the curvature associated to the $\tilde g$ metric is given by\footnote{See for instance App. D in \cite{Wald}}
\be
\tilde R=e^{-2 \phi}(R-2\Box \phi).
\en
Setting the curvature $\tilde R$ to an arbitrary constant $-\lambda$ (the sign is chosen for latter convenience) leads to the nonlinear equation 
\be
R-2\Box \phi+\lambda e^{2\phi}=0,
\label{Liouvilleeq}
\en
which is the so-called \emph{Liouville equation}. Finding a solution $\phi$ which satisfies \eqref{Liouvilleeq} is actually giving an answer to the uniformization problem. \\
Later, equation \eqref{Liouvilleeq} was interpreted by Polyakov as the equation of motion of the quantum field theory that appears in string theory when one studies how the path integral measure transforms under Weyl rescaling. The classical equation of motion of Liouville field theory would be (\ref{Liouvilleeq}), which can be derived from the \emph{Liouville action}
\be
S_{\mathrm{Liouville}}=\int d^2x \sqrt{|g|} \, 
\left(g^{ab}\p_a \phi \p_b \phi +\phi R+\frac{\lambda }{2}e^{2\phi}\right).
\label{Lcl}
\en
On the cylinder (or on the torus), one can always set the second term to zero. In this case, the Liouville action coincides, up to field redefinition, with the one obtained in \eqref{final}, which is consistent with the fact that the metric at infinity is the flat metric on the cylinder.

\subsection{At quantum level: stress tensor and central charge}
Before closing this section, let us present some quantum properties of Liouville action.
As an exact conformal field theory, the quantum Liouville action is
\begin{equation}
S_q =\frac{1}{4\pi }\int d^{2}x\sqrt{|g|}\left( g^{ab}\partial
_{a}\varphi \partial _{b}\varphi +(b+1/b)R\varphi
+4\pi \mu e^{2b\varphi }\right) ,  \label{S}
\end{equation}%
where $\mu $ is an arbitrary positive constant and $b$ is a dimensionless positive parameter which controls the quantum effects. Action \eqref{S} corresponds to a non-free scalar field theory formulated on a two-dimensional manifold doted with a metric $g_{ab}$. The curvature scalar of this two-dimensional space, $R$, couples non-minimally (linearly) with the field $\varphi$.

The value of $\mu $ can be set to $1$ without loss of generality by
shifting the field as follows $\varphi \rightarrow \varphi -(2 b)^{-1}\log \mu $. This shifting is a symmetry of the classical theory as it
merely generates a total derivative term $\propto \int d^{2}x\sqrt{|g|}R$ in the Lagrangian.

One can consistently recover the classical Liouville action by setting $\varphi=\phi/b$ and $8\pi \mu b^2 \equiv \lambda$; the action above then becomes
\begin{equation}
S_{cl}\equiv 4\pi b^2 S_q =\int_{\mathcal{C}_{2}}d^{2}x\sqrt{g}\left(g^{ab}\partial
_{a}\phi \partial _{b}\phi +R\phi(1+b^2)
+ \frac{\lambda}{2} e^{2\phi }\right),
\end{equation}%
which reduces to \eqref{Lcl} in the limit $b^2 \gt 0$ (indeed, $\hbar$ corrections correspond here to corrections of order $b^2$).

Defining $(z,\overline{z})$ the complex coordinates as usual in $2D$ CFT, one can show that the holomorphic and antiholomorphic components of the stress tensor, $T_{zz}$ and $T_{\overline{z}\overline{z}}$, are given by \cite{Zamo}
\be
\badat{2}
T &\equiv &T_{zz}=-\left( \partial \varphi \right) ^{2}+(b+1/b)\partial ^{2}\varphi , \\
\overline{T} &\equiv &T_{\overline{z}\overline{z}}=-\left( \overline{%
\partial }\varphi \right) ^{2}+(b+1/b)\overline{%
\partial }^{2}\varphi ,  \label{TTb}
\eadat
\en
with $\partial=\partial_z$, $\bar \partial=\partial_{\zb}$. 
One can compute the central charge of Liouville by computing the operator product expansion of two stress-tensor operators and, from this, one can read the central charge. More precisely, one has
\begin{equation}
T(z_{1})T(z_{2})= \frac{c_L/2}{(z_{1}-z_{2})^{4}}+ \frac{2T(z_{2})}{%
(z_{1}-z_{2})^{2}}+\frac{\partial T(z_{2})}{(z_{1}-z_{2})}+...  \label{TT}
\end{equation}%
where the ellipses stand for terms that are regular at $z_{1}=z_{2}$; which can be easily computed starting from \eqref{TTb} and using the free field correlator $\left\langle \varphi (z_{1})\varphi (z_{2})\right\rangle =- 2 \log |z_{1}-z_{2}|$. One then reads from \eqref{TT} the value of the central charge
\begin{equation}
c_L =1+6 ( b+1/b )^2.  
\label{cLiouv}
\end{equation}
Remarkably, one notices from (\ref{cLiouv}) that there is a $\mathcal{O}(1/\hbar )$ contribution to $c_L$, namely that the theory presents a classical contribution $\mathcal{O}(1/b^2 )$ to the conformal anomaly (see \cite{Jackiw} for more details).

The spectrum of primary operators in Liouville field theory is represented by the exponential vertex operators 
\begin{equation}
V_{\alpha }(z) = e^{2\alpha\phi (z)} ,
\end{equation}
which create primary states of the theory with momentum $\alpha$. The operator product expansion between the stress-tensor and these vertex operators is
\begin{equation}
T(z_1)V_{\alpha }(z_2) = \frac{\Delta}{(z_1-z_2)^2} V_{\alpha }(z_2)+ \frac{1}{(z_1-z_2)} \partial V_{\alpha }(z_2) + ... ,
\end{equation}
where the conformal dimension $\Delta $ is given in terms of the momentum by
\begin{equation}
\Delta = \alpha \left( b + \frac{1}{b} -\alpha \right)
\label{Delta} ,
\end{equation}
and analogously for the antiholomorphic component. On the other hand, normalizable states in the theory correspond to momenta
\begin{equation}
\alpha = \frac{b}{2}+\frac{1}{2b} + i s \ , \ \ \ \text{with} \ s\in \mathbb{R} . \label{LFTalpha}
\end{equation}
Therefore, the spectrum of normalizable states of Liouville field theory is \emph{continuous} and satisfies
\begin{equation}
\Delta = \frac{1}{4} \left( b + \frac{1}{b} \right)^2 +s^2 \geq \frac{1}{4} \left( b + \frac{1}{b} \right)^2  . \label{LFTspectrum}
\end{equation}
This means that the theory has a \emph{gap} between the value $\Delta = 0 $ and the minimum eigenvalue 
\be
\Delta_0 = \frac{1}{4}\left(b+\frac{1}{b}\right)^2=\frac{c_L-1}{24},
\label{minL}
\en
where the continuum starts. An observation that will be relevant later is that, in the semi-classical limit, where the central charge becomes large, this gaps reads
\begin{equation}
\Delta _0 \approx \frac{c_L}{24} . \label{LFTgap} 
\end{equation}

Before concluding this section, let us make the following comment: As mentioned in the previous sections, through the reduction from Einstein gravity to Liouville theory we did not consider the contribution of holonomies. This can be seen as a limitation since holonomies are important to describe, for instance, the BTZ solution. However, it turns out that Liouville theory {\it knows} about the holonomies. This is because the holonomies of the Chern-Simons gauge connections that correspond to the different BTZ geometries can be classified in terms of $SL(2,\mathbb{R})\times SL(2,\mathbb{R})$ conjugacy classes, and the latter are closely related to the classical solutions of Liouville field theory. While the BTZ black holes (namely $|J/\ell |\leq M>0$) correspond to the {\it hyperbolic conjugacy class} of $SL(2,\mathbb{R})$, the particle-like solutions (for example, $-1/(8G)\neq M<0$) correspond to the {\it elliptic conjugacy class} of $SL(2,\mathbb{R})$; the massless BTZ black hole ($M=J=0$) belonging to the {\it parabolic conjugacy class}. It happens that all these solutions can actually be gathered in Liouville theory by studying the monodromies of the classical solutions of the field equation\footnote{More precisely, they are related to the solutions $f$ of the so-called Hill equation $(\partial^2 + T(z))f=0$.} around singularities. We will not discuss the details of this in these notes.


\section{Liouville and the entropy of the BTZ black hole }
\label{secEnt}
In this section, we will discuss how the conformal field theory description appearing in the boundary can be used to reproduce the BTZ black hole entropy. More precisely, we will begin by reviewing how, by means of a Cardy formula, the conformal field theory structure appearing through the AdS$_3$ asymptotic symmetry manages to account for three-dimensional the black hole entropy. Then, we will discuss some issues about Liouville field theory and the BTZ black hole spectra.

\subsection{Cardy formula and effective central charge}
In a CFT$_2$, the degeneracy of states in the limit of large conformal dimension $\Delta $, and under certain assumptions, is given by the Cardy formula \cite{Cardy}. Namely, let a (chiral part of a) CFT with central charge $c$, such that its partition function on the torus is modular invariant; then, the degeneracy of states of conformal dimension $\Delta$, denoted $\rho (\Delta)$, at large $\Delta $ is given by
\be
\rho(\Delta)\approx \mathrm{exp}\left[2\pi \sqrt{\frac{c_{\mathrm{eff}}\Delta^{\text{cyl}}}{6}} \right]\rho(\Delta_0),
\label{rho}
\en
where $\Delta_0$ is the lowest eigenvalue of $L_0$ on the sphere, $\Delta^{\text{cyl}}=\Delta -c/24$ is the conformal dimension on the cylinder\footnote{The shifting $-c/24$ is related to the conformal mapping between the Riemann sphere and the cylinder, which produces a non-vanishing Schwarzian derivative in the stress-tensor transformation law.}, and $c_{\mathrm{eff}}$ is an effective central charge, defined by 
\be
c_{\mathrm{eff}}=c-24\Delta_0. \label{ceff}
\en
This result is notably relevant for the applications to three-dimensional gravity. In fact, one can show how the Cardy formula can be used to compute the entropy of BTZ black holes. This starts with the observation of Brown and Henneaux that, as we reviewed in Section \ref{secBC}, the asymptotic symmetry group of $(2+1)-$dimensional gravity with $\Lambda=-1/\ell^2$ is given by two copies of Virasoro algebra with central charges 
\be
c=\bar c=\frac{3\ell}{2G}. \label{LFTcBH}
\en

In \cite{Strominger}, Strominger made use of this result to reproduce the entropy of the BTZ black hole from the CFT$_2$ point of view. Indeed, applying Cardy formula (\ref{rho}), which can be written
\begin{equation}
S_{\text{CFT}}\equiv \log \rho (\Delta , \bar{\Delta }) = 2\pi \sqrt{\frac{c_{\mathrm{eff}} \Delta^{\text{cyl}}}{6}} + 2\pi \sqrt{\frac{\bar{c}_{\mathrm{eff}} \bar{\Delta}^{\text{cyl}}}{6}}  , \label{LFT4}
\end{equation}
using (\ref{LFTcBH}) as the effective central charge, and the fact that the BTZ conserved charges are given by 
\begin{equation}
\Delta^{\text{cyl}}=\frac{1}{2}(\ell M+J)  \ , \ \ \ \bar \Delta^{\text{cyl}}=\frac{1}{2}(\ell M-J), \label{LFTD}
\end{equation}
one finds, using relations \eqref{MJ},
\be
S_{\text{CFT}}=\frac{2\pi r_+}{4  G}=S_{\text{BH}} \label{Matching}.
\en
This is a remarkable result! It manifestly shows that the Cardy formula of the boundary CFT$_2$ exactly reproduces the entropy of the AdS$_3$ black hole \eqref{EntBH}.

In the derivation of the Cardy formula (\ref{rho}) one assumes that the conformal dimension $\Delta $ is large and the central charge $c$ finite. Notice that this is not in contradiction with the semi-classical limit (large $c$) since one can always consider large mass black holes while considering the AdS$_3$ radius much larger than the Planck length $G$. In other words, we have $\Delta \gg c \gg 1$. For different limits in relation to the Cardy formula, in which $c$ is large but not necessarily much smaller than $\Delta $, see \cite{Hartman}; see also \cite{Wittenrevisited} for a computation with small $\Delta $ (and small $c$) black holes.

\subsection{A caveat of the CFT spectrum and Liouville theory}

Let us now discuss a subtlety in relation to the Liouville field theory spectrum. Let us begin by noticing that what actually enters in the Cardy formula (\ref{LFT4}) is the \emph{effective} central charge $c_{\mathrm{eff}} $ rather than $c$. This is related to the fact that, when deriving the asymptotic growth of states at large $\Delta$, one resorts to the saddle point approximation, which selects the state of lower conformal dimension. In other words, the possibility of $c_{\mathrm{eff}} \neq c$ is associated to whether the theory has a gap or not. This is exactly what occurs in Liouville theory which, as we mentioned in (\ref{LFTspectrum}), when formulated on the sphere has a minimum eigenvalue of $\Delta_0 $ different from zero. Therefore, according to (\ref{minL}), Liouville effective central charge would be ${c_\mathrm{eff}}=c_L-24\Delta _0 = 1$, and this would be in conflict with the derivation of (\ref{Matching}), which requires $c_{\mathrm{eff}}$ to be the Brown-Henneaux central charge in order to reproduce the black hole entropy. 

This issue has been discussed in the literature \cite{CarlipCFT, CarlipWW, Martinec}, where it has been proposed that this is an indication that the description of thermodynamics in terms of Liouville field theory should be considered only as an effective description. In other words, the different steps connecting Einstein theory and Liouville theory, which were proven to hold at the level of the actions, should not be taken as a proof that the theories involved are equivalent beyond the semi-classical limit. In fact, the quantum regimes of both theories can be notably different. Liouville theory exhibits indeed many issues that make difficult to believe that it could describe three-dimensional gravity beyond the classical approximation. Nevertheless, one can still insist with the Liouville effective description and see how far it brings us. With this aim, let us discuss the Liouville theory spectrum in relation to the one of AdS$_3$ gravity in more details.

The reduction from a WZW model to Liouville can be performed at the quantum level: this is the so-called (Drinfeld-Sokolov) Hamiltonian reduction \cite{DS,BO,Balog1}. Through this procedure, one finds that the level $k=\ell/(4G)$ of the WZW action and the parameter $b$ of Liouville action are related as follows:
\be
b^2=\frac{1}{k-2},
\en
which in the semi-classical approximation $\ell \gg G$, reads $b^2\approx {1}/{k} \ll 1$.\\
Since the central charge of Liouville theory is given by \eqref{cLiouv}, one finds that, in the semi-classical approximation,
\be
c_{L}\approx 6k=\frac{3\ell}{2G}.
\en
That is, the central charge of Liouville coincides exactly with the one of Brown-Henneaux. In addition, one observes that, in the same limit, the gap in the spectrum (\ref{LFTgap}) also agrees with the gap in the spectrum of BTZ black holes. More precisely, from (\ref{LFTD}) one finds that for the BTZ black holes one has
\begin{equation}
\Delta^{\text{cyl}}+\bar{\Delta}^{\text{cyl}}=\ell M \ , \ \ \ \  \Delta^{\text{cyl}}-\bar{\Delta}^{\text{cyl}}=J \ ;
\end{equation}
on the other hand, AdS$_3$ space, which is the natural vacuum of the theory, corresponds to $\ell M=-1/(8G)$ and $J=0$, which reads
\begin{equation}
\Delta^{\text{cyl}}=\bar{\Delta}^{\text{cyl}}=-\frac{\ell}{16G} \approx  -\frac{c_L}{24} \ . \label{LFTdege}
\end{equation}
We see therefore that the mass gap of the BTZ spectrum (namely the gap $\Delta _0$ between AdS$_3$ geometry and the massless BTZ black hole) coincides with the gap in the spectrum of Liouville theory (\ref{LFTgap}). This suggests to include, apart from the normalizable states that account for the black hole spectrum, other states in Liouville theory, such as (\ref{LFTdege}). In fact, apart from the normalizable states (\ref{LFTalpha}), in Liouville field theory there exists another set of interesting operators to be explored \cite{CarlipCFT,Gaston}. These are the so-called degenerate operators, which correspond to values of momenta \cite{Zamo}
\begin{equation}
\alpha = \frac{1-n}{2b} +\frac{1-m}{2}b, \label{degerus}
\end{equation}
with $m$ and $n$ two positive integer numbers. Notice that the states (\ref{degerus}) with $m> 1$ become irrelevant in the semi-classical limit $b\to 0 $. Notice also that the state with $m=n=1$ corresponds exactly to the vacuum $\Delta =\bar{\Delta}=0$, namely $\Delta^{\text{cyl}} =\bar{\Delta}^{\text{cyl}}\approx -c/24$ as in (\ref{LFTdege}). A natural question is to find what geometries correspond to the other degenerate states, namely those with $m=1<n$ in (\ref{degerus}). According to (\ref{Delta}), in the semi-classical limit theses states must correspond to gravity solutions with charges $\Delta = \bar{\Delta } \approx (1-n^2)/(4b^2)$; that is, 
\begin{equation}
\Delta^{\text{cyl}} + \bar{\Delta }^{\text{cyl}} \approx -n^2\frac{c_L}{12} \ , \ \ \ \Delta^{\text{cyl}} + \bar{\Delta }^{\text{cyl}} =0, 
\end{equation}
or equivalently, 
\begin{equation}
\ell M = -\frac{n^2}{8G} \ , \ \ \ J=0 , \label{Degas} 
\end{equation}
with $n\in \mathbb{Z}_{>0}$. These states play an important role in the discussion \cite{Balog1} and can be shown to represent the special cases of negative mass geometries whose angular excesses around $r=0$ are integer multiples of $2\pi $. These {\it exact angular excesses} geometries also exhibit interesting supersymmetric properties \cite{BDMT1}, and their relevance in higher-spin theories in AdS$_3$ has been discussed recently \cite{Joris}. 

\section{Other directions and recent advances}
The asymptotic symmetries and dynamics of three-dimensional gravity is highly sensitive to the choice of boundary conditions. In these lectures, we have focused on the seminal Brown-Henneaux boundary conditions, where the metric at the boundary has no dynamics. In the last years, there have been many works generalizing or modifying these boundary conditions. In \cite{CompereMarolf}, the boundary metric is allowed to be dynamical; in \cite{CSS1}, chiral boundary conditions were given, while in \cite{CedricBC,IndBC} the boundary metric is in a conformal gauge and light-cone gauge, respectively. These new notions of asymptotically AdS$_3$ spacetimes provide new potential CFT$_2$s living at the boundary \cite{CSS2,Porrati,GBH}. 

Also, the scope of these lecture notes was limited to the case of pure gravity, but the study of asymptotic symmetries and dynamics of three-dimensional AdS spaces similar to the one addressed here can be extended to broader set-ups such as supergravity \cite{Banados1,Maoz} and higher-spin theories \cite{HS1,HS2}. 

Before concluding these notes, let us mention an important direction that these notes have not explored, since, motivated by AdS/CFT, we focused on the case of Anti-de Sitter spacetimes. However, over the last years, significant effort has been made to extend holographic tools to the case of non-AdS spacetimes. In particular, holographic properties of gravitational theories with a vanishing cosmological constant have been investigated. In this flat holography perspective, a rich asymptotic dynamics can be found at null infinity: the symmetry algebra of asymptotically flat spacetimes is the BMS$_3$ algebra \cite{bms31, bms32}
\be
\badat{3}
&i\{J_m,J_n\}=(m-n)J_{m+n}+\frac{c_1}{12}m^3 \delta_{m+n,0}, \\
&i\{P_m,P_n\}=0,\\
&i\{J_m,P_n\}=(m-n) P_{m+n}+\frac{c_2}{12}m^3 \delta_{m+n,0},
\label{bms3}
\eadat
\en
with $c_1=0$, $c_2=3/G$. Algebra \eqref{bms3} is an infinite-dimensional algebra made out of supertranslations and superrotations generated by $P_m$ and $J_m$, respectively. This algebra can be obtained from the (two copies of the) Virasoro algebra \eqref{Vir} by writing the latter in terms of the generators $P_m=(L_m+\bar L_{-m})/\ell$, $J_m=(L_m-\bar L_{-m})$ and then taking the limit $\ell \gt \infty$. One can also connect through a well-defined flat-space limit \cite{Hernan1} the phase space of asymptotically AdS and flat spacetimes: the limit of the BTZ black holes are cosmological solutions whose thermodynamical properties can be understood from a holographic perspective \cite{Cardyflat1,Cardyflat2}. Finally, let us mention the fact that the dual dynamics of three-dimensional asymptotically flat spacetimes at null infinity has been shown to be a BMS$_3$ invariant Liouville theory \cite{Hernan2, Hernan3} (generalized afterward adding higher-spin fields \cite{Pino} and supersymmetry \cite{BDMT2}), through a reduction very similar to the one presented here; one goes first through a chiral iso$(2,1)$ WZW model and the Hamiltonian reduction reduces further the theory to a flat chiral boson action which can be finally related to a Liouville theory.

\section*{Acknowledgments}
\appendix
I would like to thank all the organizers of this great edition 2015 of the Modave Summer School, as well as all the participants for their questions, comments and interesting discussions we shared during the lectures. I also want to thank Glenn Barnich, Gaston Giribet, and Hern\'an A. Gonz\'alez for enlightening discussions on these topics.  L.D.~is a FRIA research fellow of the FNRS Belgium.

\section{Conventions}
\label{conventions}
Our conventions are such that the Levy-Civita symbol fulfills $\epsilon
_{012}=1$, and the tangent space metric $\eta_{ab}$, with $a=0,1,2$, is
off-diagonal, given by%
\[
\eta_{ab}=\left(
\begin{array}
[c]{ccc}%
0 & 1 & 0\\
1 & 0 & 0\\
0 & 0 & 1
\end{array}
\right)  \ .
\]
As $sl(2, \mathbb R)$ generators, we take
\[
j_{0}=\frac{1}{\sqrt 2}\left(
\begin{array}
[c]{cc}%
0 & 1\\
0 & 0
\end{array}
\right)  \virg j_1=\frac{1}{\sqrt 2}\left(
\begin{array}
[c]{cc}%
0 & 0\\
1 & 0
\end{array}
\right)  \virg  j_2=\frac{1}{2}\left(
\begin{array}
[c]{cc}%
1 & 0\\
0 & -1
\end{array}
\right)  \ ,
\]
which satisfy $ [j_a,j_b]=\ep_{abc} j^c $, Tr$(j_a j_b)=\frac{1}{2}\eta_{ab}$ ($a,b =0,1,2$). \\

For the two-dimensional Wess-Zumino-Witten model, we use the coordinates $x^0=\tau, x^1=\varphi$ ($\mu, \nu= 0,1$), with $\eta_{\mu \nu}=\mathrm{diag}(-1,1)$ and $\ep^{\mu \nu}$ with $\ep^{01}=1$. In light-cone coordinates $x^\pm=\tau \pm \varphi $, we have $\eta^{+-}=-2=\eta^{-+}$, $\eta^{++}=0=\eta^{--}$, and $\ep^{+-}=-2$.

\def\cprime{$'$}

\pagestyle{empty}
\providecommand{\href}[2]{#2}\begingroup\raggedright\endgroup

\end{document}